# A Quaternion Based Quantum Chemical *ab initio* Treatment of Coherent and Non-Coherent Electron Transport in Molecules

Augusto C. L. Moreira and Celso P. de Melo*

Departamento de Física, Universidade Federal de Pernambuco

50670-901 Recife, PE, Brazil

## Abstract

We present a quaternion inspired formalism specifically developed to evaluate the intensity of the electrical current that traverses a single molecule connected to two semi-infinite electrodes as the applied external bias is varied. The self-adjustment of the molecular levels is fully described at a density functional *ab initio* quantum chemical level. Use of a quaternion approach allows for an integrated treatment of both coherent (ballistic) and non-coherent (co-tunneling) contributions to the effective charge transport, where the latter involve the existence of transient charged states of the corresponding molecular species. An expression for the net current is calculated by using second-order perturbation theory to take into account all possible transitions between states localized at the two different electrodes that involve intermediary levels in the so-called "extended molecule" complex that comprises the system of interest attached to two small metallic clusters. We show that by a judicious choice of the relevant molecular parameters, the formalism can be extended to describe the electronic transport both in conjugated as in saturated molecules, where localized orbitals are more likely to be found. In this manner, the method can be applied to the full range of coupling regimes, not only to the weak or strong cases, but also in intermediate situations, where ballistic and co-tunneling processes may coexist.

# 1. Introduction

## *1.1. Conductance through single–molecules*

In their landmark work [1], Aviram and Ratner first proposed the concept of a molecular rectifier, and since then a crescent amount of theoretical effort has been dedicated to the understanding of electronic transport processes occurring in the molecular scale [2]. However, after so many years of hard and varied effort, the calculation of the current that flows through a single molecule connected to metal electrodes under an externally applied bias voltage remains one of the most difficult theoretical problems to be properly considered. In fact, when dealing with structures which are much smaller than the electron mean free path, one must reformulate concepts like resistance and capacitance. For instance, a well-known result for large conductors is that the conductance is proportional to the inverse of a characteristic length ($G \sim L^{-1}$) [2], so that an increasing conductance should be expected as the typical size of the device is progressively reduced. However, experiments actually show that the measured conductance tends to a limiting value $G_{lim}$ when the length of the conductor becomes much shorter than the electron mean free path [3]. A first theoretical insight into the question appeared with Landauer's introduction of the concept of 'conductance from transmission' [4, 5], in which the resistance of a very small conductor is associated to the "coherent" probability of transmission ($T$) through the "internal" transversal modes, supposed to be limited in number when compared to the infinite range of possible modes in the external electrodes.

## *1.2. Coherent and non-coherent transport*

Different suggestions of how to calculate $T$ have been introduced, the most successful of them based on the use of Green´s function techniques, such as the Non Equilibrium Green´s Function (NEGF – also known as the Keldysh formalism) [6] and the recursive Green´s function (RGF) method [3]. In the former type of approach, the molecular coupling to the two terminal contacts (which are labeled as $L$ and $R$, respectively) is described by an effective device Hamiltonian $H^{eff} \rightarrow H + \Sigma_L + \Sigma_R$ that incorporate each electrode by taking into account its corresponding self-energy matrix $\Sigma_i$ (i = $L$, $R$). Once this effective Hamiltonian is known, the overall current through the system can be calculated and, as long as its coherence is preserved, the electron transport maintains a "ballistic" character with no evidence of scattering from the molecular internal structure, a phenomenon which would be typically identified to the appearance of spatially localized molecular states [4]. The occurrence of ballistic transport is associated to the existence of a strong interaction between the molecule and the metallic leads, in the so called strong coupling regime. Within the Keldysh formalism, the transmission function through the system can be written as $T(E) = \Gamma_L G \Gamma_R G^{+}$ [6], where $G$ is the Green's function of the coupled system and $\Gamma_i$ is the broadening contribution of the molecular levels (which is associated to the anti-hermitian part of $\Sigma_i$). The electrical current in this case is given by

$$I = \frac{2e}{h} \int_{-\infty}^{\infty} \bar{T}(E) \left[ f_{(E-\mu_1)} - f_{(E-\mu_2)} \right] dE \quad , \qquad (1)$$

where $f_{(E-\mu_i)}$ $(i=1,2)$ are the Fermi distribution functions at the two semi-infinite electrodes.

One also uses Green's functions in the RGF procedure, although in a recursive manner. For this, after the region of the device is discretized, the total Green´s function $G$ can be obtained by use of Dyson's equation, $G = G_0 + G_0VG$, where $G_0$ is the unperturbed Green´s function and $V$ a small perturbation. In this case, the transmission function is given by $T(E) = V + VGV$ [7].

It is worthwhile to note that even though the methods mentioned above are strictly valid only for the case of coherent transport, incoherent phenomena such as electron-phonon coupling can be described by adding an interaction potential and then evaluating the corresponding coupling constants. In fact, as shown by Datta [4], while in the case of NEGF method this procedure would be equivalent to attach an extra "contact" to the device with a self-energy ($\Sigma_S$) and associated to an additional broadening ($\Gamma_S$) of the involved states, in the RGF method one makes the replacement $V \to V + U$. In both cases, the complete wave function must now take into account both (phonon and electronic) individual sub-spaces.

In the usual treatment of molecular electronic devices [8-10], even after inclusion of non-coherent contributions the transport through the system remains confined to a single electronic potential surface, and the active part of the system (a single molecule for example) must remain unaltered (i.e., in the neutral form or in a specific anionic or cationic state). However, this picture is not valid if the coupling between the molecule and the metallic leads is sufficiently small; in this limit, the charge transport is sequential and we must allow for the presence of transient charged states of the molecular species considered [6]. One faces now the regime of weak coupling, where charging effects such as Coulomb blockade [3], for example, may occur.

In fact, while methods such as NEGF are not able to describe charge flow in a correct manner, a more complete description of quantum transport through a single molecule connected to two terminal electrodes has to account for the occurrence of not only the two limiting regimes previously described, but also the intermediate case [6], where the multi electron master equation (MEME) approach does not include broadening effects properly. In the intermediate regime, accordingly to the intensity of the externally applied electric field, either ballistic processes or alternative charge transfer pathways involving transient charged molecular states (some of them corresponding to non-coherent tunneling events at the two electrode-molecule junctions [11]) may dominate the overall charge transport.

### *1.3. The extended molecule concept*

In the present paper we will develop a new formalism for the problem by considering a model system in which a single molecule (*M*) is connected to two identical clusters ($C_L$ and $C_R$) of metal (hereafter, gold) atoms that will mimic the terminal ends of semi-infinite circuit leads. This will form the "extended molecule" (*EM*) whose electronic structure will be calculated anew at the *ab initio* level at each chosen value of the external potential bias. Finally, the $EM \equiv C_L - M - C_R$ will be coupled to the left and right electrodes (*LE* and *RE*, respectively) so that the overall electronic current traversing the molecule can be determined for the entire range of applied potentials (see Fig. 1).

It is especially important to note at this point that it is not a straightforward issue to decide *a priori* if a given molecular system will be strongly or weakly coupled to a pair of existing metallic electrodes. A criterion usually adopted [6] is to assume that a strong coupling will be developed through the molecular system if the corresponding molecular orbitals are sufficiently delocalized to present nonvanishing electronic densities at the opposite sides of the extended molecule [12]. (Also note that this assumption precludes the existence of strong coupling – and therefore of the occurrence of ballistic transport – in saturated bridge type of molecules, where spatial localization of the electronic density at each extremity usually occurs.)

However, a logical quagmire rapidly develops if one stretches the argument to consider larger and larger molecular entities: while, on one hand, a progressive increase in the number of metallic atoms in the terminal clusters of the EM should improve the quality of the calculated results [13], on the other hand, the larger the size of the EM, less likely it becomes to find occupied states that are truly delocalized throughout the extended molecule region. But, this tendency of spatial localization in large systems is exactly at the origin to the so-called "problem of bound states", since no unique expression can be derived for the density matrix of electronic states that do not diffuse to the semi-infinite electrodes region [14]. To compound to the difficulties, if the theoretical treatment of the problem allows for a new *ab initio* calculation of the molecular system every time that the value of the external electric field is adjusted, the spatial localization of the frontier molecular orbitals can change accordingly, and consequently, the nature of the transport regime would vary along the actual calculation of the overall current.

The above limitations are avoided in the present formalism, since it was developed to account not only for the case of coherent (ballistic) transport, but also to allow for the inclusion of the alternative mechanisms where – under a fixed externally applied bias – the initial step in the process involves the transfer of one electron, either one moving from the cathode to the extended molecule or another one flowing from the EM to the anode; in either one of these cases, it will be assumed that the electronic distribution of the newly charged molecular species will adjust itself to the novel situation before the electron transport between the electrodes is complete, so that the quantum-chemical problem to be solved would be that of the corresponding anion or cation (see Scheme A, where we assume a positive bias, i.e., the *LE* [*RE*] is the cathode [anode]). As equivalent pictures, these two non-coherent ("co-tunneling") processes can be thought as corresponding to the injection of one extra electron into an unoccupied molecular orbital (UMO) or the introduction of a "hole" into one of the occupied molecular orbitals (OMO) of the $EM$ .

Therefore, these co-tunneling contributions to the total current in the device differ fundamentally from the ballistic mode by the fact that they correspond to physical situations where transitions between different potential surfaces occur; note that for this, two non-correlated charge transfer steps must be sequentially involved, so that during the intermediate time interval a transient charged molecular state temporarily flicks into existence. While only a limited number of proposals have considered this manifold of possibilities [11, 15], in the present work we will develop a unified treatment that will allow the extended molecule to be described at a density functional quantum chemical level of treatment so that the relative contributions of the coherent and incoherent regimes could be calculated. The former is expected to be dominant in conjugated molecules – see Ref. [16], while the latter becomes more relevant in the case of saturated systems (Ref. [17]).

## 2. The quaternionic formalism

*2.1. Molecular charge states as independent quaternionic sub-spaces*

We will now introduce a new formalism for obtaining the electrical current that crosses the extended molecule comprising a single organic molecule of interest attached to two small metallic (gold) clusters, which are in turn connected to infinite leads through which a transverse electric field of varying intensity is applied. For a fixed value of the applied potential, once the three possible pathways depicted in Scheme A are independently considered, both ballistic and co-tunneling terms will arise naturally in the resulting expression for the overall electrical current. To be able to do this in a simple and elegant manner, we will recur to the use of a *quaternion*-inspired formalism. Quaternions is a theoretical development originally introduced by W. R. Hamilton in the XIX Century [18] that in modern times has been adapted by Adler [19] to quantum mechanical problems.

Here we use the notion of quaternions in an novel manner in that our main *ansatz* is to attribute for each different charge state of the molecule (labeled by *1,2* and *3*) a distinct imaginary unit (*i*, *j*, and *k,* such that they obey the commutation rules $ij=-ji=k$ and its cyclic permutations) and construct a more complete system composed of three uncoupled time dependent Schrodinger equations, each one belonging to a different quaternionic sub-space specific to one of these possible molecular charge states, in the form

$$
\begin{aligned}
&i\hbar\frac{\partial\psi_{n_1}(r,t)}{\partial t}=H_1\psi_{n_1}(r,t) &&\text{(quantum states of the neutral molecular species)}\\
&j\hbar\frac{\partial\psi_{n_2}(r,t)}{\partial t}=H_2\psi_{n_2}(r,t) &&\text{(quantum states of the anionic species)}\\
&k\hbar\frac{\partial\psi_{n_3}(r,t)}{\partial t}=H_3\psi_{n_3}(r,t) &&\text{(quantum states of the cationic species)}\ .
\end{aligned}
\tag{2}
$$

(Note that, for any of those charge states – let's say, *x=1*, for example – the corresponding zeroth-order Hamiltonian can be considered as the direct sum

$$
H_1=H_1^{LE}\oplus H_1^{EM}\oplus H_1^{RE} \quad , \tag{3}
$$

since the semi-infinite electrodes and the extended molecule are assumed not to be initially coupled.)

Alternatively, Eq. (2) can be written in the matrix form

$$
\begin{pmatrix} i\hbar\frac{\partial}{\partial t} & 0 & 0\\ 0 & j\hbar\frac{\partial}{\partial t} & 0\\ 0 & 0 & k\hbar\frac{\partial}{\partial t}\end{pmatrix}\begin{pmatrix}\psi_{n_1}(r,t)\\ \psi_{n_2}(r,t)\\ \psi_{n_3}(r,t)\end{pmatrix}=\begin{pmatrix}H_1 & 0 & 0\\ 0 & H_2 & 0\\ 0 & 0 & H_3\end{pmatrix}\begin{pmatrix}\psi_{n_1}(r,t)\\ \psi_{n_2}(r,t)\\ \psi_{n_3}(r,t)\end{pmatrix} \quad , \tag{4}
$$

where the time dependent part of each one of the three quaternionic wavefunctions is given by the usual exponential function associated to its respective complex number, i.e.

$$\begin{pmatrix} \psi_{n_1}(r,t) \\ \psi_{n_2}(r,t) \\ \psi_{n_3}(r,t) \end{pmatrix} = \begin{pmatrix} \psi_{n_1(r)}.e^{\frac{-iE_{n_1}t}{\hbar}} \\ \psi_{n_2(r)}.e^{\frac{-jE_{n_2}t}{\hbar}} \\ \psi_{n_3(r)}.e^{\frac{-kE_{n_3}t}{\hbar}} \end{pmatrix} . \tag{5}$$

By this construct, each one of these wavefunctions is automatically orthonormalized, i.e., $\left\langle \psi_{n_x}(r,t) \middle| \psi_{m_y}(r,t) \right\rangle = \delta_{x,y}\delta_{n_x,m_y}$, where $x$ and $y$ denote the charge states, i.e., $x$, $y$ = 1,2 or 3 for the neutral form of the *EM*, and its anion and cation varieties, respectively.

Insofar its unfamiliar form, Eq. (4) corresponds to just a system composed by three uncoupled time dependent Schrodinger equations (each one corresponding to a different quaternionic subspace represented by a distinct imaginary part) that can be separately solved.

*2.2. The coupling of the "extended molecule"*

We will assume that the molecule of interest is terminally coupled to two gold clusters forming an extended-molecule (*EM*) connected to the left and right metallic electrodes (*LE* and *RE*, respectively), and that the solutions for each one of the three independent components is known in the form

$$H_x^{LE}\left|\psi_{m_x}^{LE}\right\rangle = E_{m_x}^{LE}\left|\psi_{m_x}^{LE}\right\rangle \quad ,$$
$$H_x^{EM}\left|\psi_{l_x}^{EM}\right\rangle = E_{l_x}^{EM}\left|\psi_{l_x}^{EM}\right\rangle \tag{6}$$
and
$$H_x^{RE}\left|\psi_{n_x}^{RE}\right\rangle = E_{n_x}^{RE}\left|\psi_{n_x}^{RE}\right\rangle \quad ,$$

where $H_{LE}$ and $H_{RE}$ are the corresponding Hamiltonians of two semi-infinite electrodes, and $H_{EM}$ is the Hamiltonian of the extended-molecule, which will be here determined at the density functional level of *ab initio* theory. Note that, in agreement with Eq. (3), if the 3-part system is initially considered to be uncoupled, the zero order states for a particular quaternionic sub-space will be direct product of states describing each individual sub-system, i.e. $\left|\psi_x\right\rangle = \left|\psi_x^{LE}\right\rangle \otimes \left|\psi_x^{EM}\right\rangle \otimes \left|\psi_x^{RE}\right\rangle$.

If a perturbation $T$ is now switched on, the complete Hamiltonian for a particular charge subspace x (i.e., $x=1$ for the neutral molecule, $x=2$ for the single-charged anion, $x=3$ for the single-charged cation) of the extended molecule, for example, will be written as

$$H_x = H_x^{LE} + H_x^{RE} + H_x^{EM} + T_x \ . \tag{7}$$

In the Transfer Hamiltonian approach [3], for example, $T$ represents the 'tunneling' of an electron from an occupied level of an electrode to an unoccupied molecular level, or vice versa. If the perturbation $T$ is sufficiently small when compared to the typical energies of the system, a time-dependent perturbation formalism approach is valid and the total Hamiltonian for a particular x-charge state of the system can be written in the usual way as

$$H_x = \sum_{n_x} E_{LE,n_x} a^{\dagger}_{LE,n_x} a_{LE,n_x} + \sum_{l_x} E_{EM,l_x} a^{\dagger}_{EM,l_x} a_{EM,l_x} + \sum_{m_x} E_{RE,m_x} a^{\dagger}_{RE,m_x} a_{RE,m_x} +$$

$$+ \sum_{m_x,l_x} T_{m_x,l_x} a^{\dagger}_{RE,m_x} a_{EM,l_x} + \sum_{l_x,n_x} T_{l_x,n_x} a^{\dagger}_{EM,l_x} a_{LE,n_x} + cc\,, \tag{8}$$

where $a^{\dagger}_{\sigma,m_x}$ and $a_{\sigma,m_x}$ are creation and annihilation Fermion operators and the subscripts $\sigma$ = *LE*, *RE* and *EM* refer to the left electrode, the right electrode and the extended molecule, respectively. One electron can only be transferred from the cathode to the extended molecule or from the extended molecule to the anode by moving from one occupied level to an unoccupied one, as properly indicated in the above expression. Note that up to now no change is admitted in the total charge state of the molecule. However, processes may occur where the time lag between the consecutive electron (or hole) transfers allows the formation of a single charged cation or anion. We will now show that the quaternionic formalism can incorporate these two additional possibilities in a very expedite manner.

*2.3. Mixing of different quaternionic sub-spaces*

Now suppose that a perturbation that mixes these different subspaces is introduced in the problem, such that if the system is in a particular charge state ($\psi_{n_1}(r,t)$, for example), it could evolve in three alternative manners, either by preserving the same charge, or by changing to different charge states ($\psi_{n_2}(r,t)$ or $\psi_{n_3}(r,t)$). At this moment, it is convenient to introduce [20] a generalized Quaternionic Operator ($Q_\alpha(t)$) capable of describing the mixing of the three quaternionic subspaces considered in Scheme A. To do this let's first define a normalized quaternionic wave function $|\chi_\alpha(t)\rangle$ as

$$|\chi_\alpha(t)\rangle = \begin{pmatrix} \gamma_1 |\psi_{n_1}(t)\rangle \\ \gamma_2 |\psi_{n_2}(t)\rangle \\ \gamma_3 |\psi_{n_3}(t)\rangle \end{pmatrix} , \tag{9}$$

where $\gamma_n$ are (normalization) constants whose properties and physical meaning will be discussed latter, so that the Quaternionic Operator ($Q_\alpha(t)$ ) would be given by

$$Q_\alpha(t) = |\chi_\alpha(t)\rangle\langle\chi_\alpha(t)| =$$

$$= \begin{pmatrix} \gamma_1 \left|\psi_{n_1}(t)\right\rangle\left\langle\psi_{n_1}(t)\right| \gamma_1 & \gamma_1 \left|\psi_{n_1}(t)\right\rangle\left\langle\psi_{n_2}(t)\right| \gamma_2 & \gamma_1 \left|\psi_{n_1}(t)\right\rangle\left\langle\psi_{n_3}(t)\right| \gamma_3 \\ \gamma_2 \left|\psi_{n_2}(t)\right\rangle\left\langle\psi_{n_1}(t)\right| \gamma_1 & \gamma_2 \left|\psi_{n_2}(t)\right\rangle\left\langle\psi_{n_2}(t)\right| \gamma_2 & \gamma_2 \left|\psi_{n_2}(t)\right\rangle\left\langle\psi_{n_3}(t)\right| \gamma_3 \\ \gamma_3 \left|\psi_{n_3}(t)\right\rangle\left\langle\psi_{n_1}(t)\right| \gamma_1 & \gamma_3 \left|\psi_{n_3}(t)\right\rangle\left\langle\psi_{n_2}(t)\right| \gamma_2 & \gamma_3 \left|\psi_{n_3}(t)\right\rangle\left\langle\psi_{n_3}(t)\right| \gamma_3 \end{pmatrix} . \qquad (10)$$

One can easily verify that $Q$ is idempotent. Please note that in the above expression: *i)* we must take into account the quaternionic non-commutative property of the different imaginary units in the time-dependent parts of the wavefunctions for the $Q_{(t)\alpha}$ operator (and so, for instance, $i.e^{-j\frac{E}{\hbar}t} \neq e^{-j\frac{E}{\hbar}t}.i$ and $e^{-i\frac{E_1}{\hbar}t} e^{-j\frac{E_2}{\hbar}t} \neq e^{-j\frac{E_2}{\hbar}t}.e^{-i\frac{E_1}{\hbar}t}$), *ii)* the off-diagonal terms in Eq.(10) induce the "mixing" of the different charge states, so that $Q_\alpha(t)$ can be interpreted as a kind of a quaternionic Hubbard operator (as opposite to a "regular" Hubbard operator that would only "mix" different states within the same quaternionic charge manifold), and *iii)* if the state $\left|\chi_\alpha(t)\right\rangle$ is normalized, then we must have $\gamma_1^2 + \gamma_2^2 + \gamma_3^2 = 1$. Strictly speaking, the $Q$-Operator ($Q_\alpha(t)$) – defined above – assures us that, if the quaternionic state $\left|\chi_\alpha(t)\right\rangle$ is normalized, we must have a probability $\gamma_l^2$ in finding this state in the $l$-th charge state. Thereby, if one has the case where the three charge states are on equal footing in terms of probability, then $\gamma_1 = \gamma_2 = \gamma_3 = 1/\sqrt{3}$ and the corresponding quaternionic operator would be written as

$$Q_\alpha(t) = \frac{1}{3}\begin{pmatrix} \left|\psi_{n_1}(t)\right\rangle\left\langle\psi_{n_1}(t)\right| & \left|\psi_{n_1}(t)\right\rangle\left\langle\psi_{n_2}(t)\right| & \left|\psi_{n_1}(t)\right\rangle\left\langle\psi_{n_3}(t)\right| \\ \left|\psi_{n_2}(t)\right\rangle\left\langle\psi_{n_1}(t)\right| & \left|\psi_{n_2}(t)\right\rangle\left\langle\psi_{n_2}(t)\right| & \left|\psi_{n_2}(t)\right\rangle\left\langle\psi_{n_3}(t)\right| \\ \left|\psi_{n_3}(t)\right\rangle\left\langle\psi_{n_1}(t)\right| & \left|\psi_{n_3}(t)\right\rangle\left\langle\psi_{n_2}(t)\right| & \left|\psi_{n_3}(t)\right\rangle\left\langle\psi_{n_3}(t)\right| \end{pmatrix} , \qquad (11)$$

and so on. Hence, a preliminary step is to ponder about the choice of values of $\gamma$ most appropriate to each specific "physical situation", as we will discuss further below.

For the moment, in the present case pictorially represented by Scheme A we will define $\eta_{x,y} = \gamma_x\gamma_y$ (with *x, y = 1, 2, 3*) such that $\eta_{x,y}^2$ is the (phenomenological) probability that a state labeled by $x$ may be "transformed" into a state labeled by $y$. Once we have constructed the quaternionic operator, we can now consider the presence of a perturbation. The usual way to do it [20, 21] would be to write, for example, $V_2(t)X(t)_{n_1,n_2}c_{n_2}$, where *X(t)* is the usual Hubbard operator. In dealing with multi-charge states, however, we have found more convenient to write the perturbation in a matrix form that allows for the different possible molecular charge states. We can define a matrix *T(t)* as

$$T(t) = \begin{pmatrix} T_1(t) & 0 & 0 \\ 0 & T_2(t) & 0 \\ 0 & 0 & T_3(t) \end{pmatrix} , \qquad (12)$$

so that, after applying the $Q$-Operator, one gets

$$Q_\alpha(t)T(t) = \begin{pmatrix} \gamma_1\gamma_1 \left|\psi_{n_1}(t)\right\rangle\left\langle\psi_{n_1}(t)\right|T_1(t) & \gamma_1\gamma_2 \left|\psi_{n_1}(t)\right\rangle\left\langle\psi_{n_2}(t)\right|T_2(t) & \gamma_1\gamma_3 \left|\psi_{n_1}(t)\right\rangle\left\langle\psi_{n_3}(t)\right|T_3(t) \\ \gamma_2\gamma_1 \left|\psi_{n_2}(t)\right\rangle\left\langle\psi_{n_1}(t)\right|T_1(t) & \gamma_2\gamma_2 \left|\psi_{n_2}(t)\right\rangle\left\langle\psi_{n_2}(t)\right|T_2(t) & \gamma_2\gamma_3 \left|\psi_{n_2}(t)\right\rangle\left\langle\psi_{n_3}(t)\right|T_3(t) \\ \gamma_3\gamma_1 \left|\psi_{n_3}(t)\right\rangle\left\langle\psi_{n_1}(t)\right|T_1(t) & \gamma_3\gamma_2 \left|\psi_{n_3}(t)\right\rangle\left\langle\psi_{n_2}(t)\right|T_2(t) & \gamma_3\gamma_3 \left|\psi_{n_3}(t)\right\rangle\left\langle\psi_{n_3}(t)\right|T_3(t) \end{pmatrix}. \tag{13}$$

*2.4. Time-evolution of the different charged molecular states*

If we now add this perturbation to the Hamiltonian of Eq. (2) we have

$$\begin{pmatrix} i\hbar\frac{\partial}{\partial t} & 0 & 0 \\ 0 & j\hbar\frac{\partial}{\partial t} & 0 \\ 0 & 0 & k\hbar\frac{\partial}{\partial t} \end{pmatrix}\begin{pmatrix} \gamma_1\left|\psi_1(t)\right\rangle \\ \gamma_2\left|\psi_2(t)\right\rangle \\ \gamma_3\left|\psi_3(t)\right\rangle \end{pmatrix} = \\ = \begin{pmatrix} H_1 & 0 & 0 \\ 0 & H_2 & 0 \\ 0 & 0 & H_3 \end{pmatrix} + Q_\alpha(t)\begin{pmatrix} T_1(t) & 0 & 0 \\ 0 & T_2(t) & 0 \\ 0 & 0 & T_3(t) \end{pmatrix}\begin{pmatrix} \gamma_1\left|\psi_1(t)\right\rangle \\ \gamma_2\left|\psi_2(t)\right\rangle \\ \gamma_3\left|\psi_3(t)\right\rangle \end{pmatrix}, \tag{14}$$

and the three Schrodinger equations become coupled, so that the usual time dependent perturbation theory will allow a quaternionic charge state of one kind to evolve to a different one (when the off-diagonal terms of $Q$ will be the determinant ones) or keep the same charge state (when the diagonal terms of $Q$ will play the most important role). Hence, the wave function will now be described by a linear combination of these states whose coefficients evolve as function of time as

$$\begin{pmatrix} \gamma_1\left|\psi_1(t)\right\rangle \\ \gamma_2\left|\psi_2(t)\right\rangle \\ \gamma_3\left|\psi_3(t)\right\rangle \end{pmatrix} = \begin{pmatrix} \gamma_1\sum_{n_1} C_{n_1}(t)\left|\psi_{n_1}\right\rangle.e^{\frac{-iE_{n_1}t}{\hbar}} \\ \gamma_2\sum_{n_2} C_{n_2}(t)\left|\psi_{n_2}\right\rangle.e^{\frac{-jE_{n_2}t}{\hbar}} \\ \gamma_3\sum_{n_3} C_{n_3}(t)\left|\psi_{n_3}\right\rangle.e^{\frac{-kE_{n_3}t}{\hbar}} \end{pmatrix}. \tag{15}$$

At this moment it is convenient to mention that the above treatment can be considered as a modified version of the approximation pioneered by Bardeen in his many-particle treatment of the tunneling problem [22]: in this so-called Bardeen's approach [23], the initial quantum state associated to the Hamiltonian of the problem has its time evolution determined by a related but distinct Hamiltonian. In the approximated ''single particle'' version of Bardeen's approach [24], the time-evolution of the total quantum state will be a combination of the independent time-evolution of each possible initial state considered.

From Eq. (15), the rate of evolution of each independent charge state can be determined by solving the differential equations

$$
\begin{pmatrix} \gamma_1 \dfrac{dC_{m_1}(t)}{dt} \\ \gamma_2 \dfrac{dC_{m_2}(t)}{dt} \\ \gamma_3 \dfrac{dC_{m_3}(t)}{dt} \end{pmatrix} =
\begin{bmatrix}
\dfrac{-i\gamma_1}{\hbar} \begin{pmatrix} \sum_{n_1} \gamma_1^2 T_{m_1,n_1}(t) e^{i\omega_{m_1,n_1} t} C_{n_1}(t) + \sum_{n_2} \gamma_2^2 T_{m_2,n_2}(t) e^{j\omega_{m_2,n_2} t} C_{n_2}(t) + \\ \sum_{n_3} \gamma_3^2 T_{m_3,n_3}(t) e^{k\omega_{m_3,n_3} t} C_{n_3}(t) \end{pmatrix} \\
\dfrac{-j\gamma_2}{\hbar} \begin{pmatrix} \sum_{n_1} \gamma_1^2 T_{m_1,n_1}(t) e^{i\omega_{m_1,n_1} t} C_{n_1}(t) + \sum_{n_2} \gamma_2^2 T_{m_2,n_2}(t) e^{j\omega_{m_2,n_2} t} C_{n_2}(t) + \\ \sum_{n_3} \gamma_3^2 T_{m_3,n_3}(t) e^{k\omega_{m_3,n_3} t} C_{n_3}(t) \end{pmatrix} \\
\dfrac{-k\gamma_3}{\hbar} \begin{pmatrix} \sum_{n_1} \gamma_1^2 T_{m_1,n_1}(t) e^{i\omega_{m_1,n_1} t} C_{n_1}(t) + \sum_{n_2} \gamma_2^2 T_{m_2,n_2}(t) e^{j\omega_{m_2,n_2} t} C_{n_2}(t) + \\ \sum_{n_3} \gamma_3^2 T_{m_3,n_3}(t) e^{k\omega_{m_3,n_3} t} C_{n_3}(t) \end{pmatrix}
\end{bmatrix} \qquad (16)
$$

where $T_{m_x,n_x}(t) = \left\langle \psi_{m_x} \middle| T_x(t) \middle| \psi_{n_x} \right\rangle$ and $\omega_{m_x,n_x} = \left(E_{m_x} - E_{n_x}\right)\hbar^{-1}$ with $x$ = *1, 2, 3.* After integration, the above expression can be written as

$$\begin{pmatrix} \gamma_1 C_{m_1}^{(n)}(t) \\ \gamma_2 C_{m_2}^{(n)}(t) \\ \gamma_3 C_{m_3}^{(n)}(t) \end{pmatrix} = \begin{pmatrix} \gamma_1 \int_{t_0}^{t} dt' \left[ \frac{-i}{\hbar} \sum_{n_1} \gamma_1^2 T_{m_1,n_1}(t') e^{i\omega_{m_1,n_1} t'} C_{n_1}^{(n-1)}(t') - \frac{i}{\hbar} \sum_{n_2} \gamma_2^2 T_{m_2,n_2}(t') e^{j\omega_{m_2,n_2} t'} C_{n_2}^{(n-1)}(t') \right. \\ \left. - \frac{i}{\hbar} \sum_{n_3} \gamma_3^2 T_{m_3,n_3}(t') e^{k\omega_{m_3,n_3} t'} C_{n_3}^{(n-1)}(t') \right] \\ \gamma_2 \int_{t_0}^{t} dt' \left[ \frac{-j}{\hbar} \sum_{n_1} \gamma_1^2 T_{m_1,n_1}(t') e^{i\omega_{m_1,n_1} t'} C_{n_1}^{(n-1)}(t') - \frac{j}{\hbar} \sum_{n_2} \gamma_2^2 T_{m_2,n_2}(t') e^{j\omega_{m_2,n_2} t'} C_{n_2}^{(n-1)}(t') \right. \\ \left. - \frac{j}{\hbar} \sum_{n_3} \gamma_3^2 T_{m_3,n_3}(t') e^{k\omega_{m_3,n_3} t'} C_{n_3}^{(n-1)}(t') \right] \\ \gamma_3 \int_{t_0}^{t} dt' \left[ \frac{-k}{\hbar} \sum_{n_1} \gamma_1^2 T_{m_1,n_1}(t') e^{i\omega_{m_1,n_1} t'} C_{n_1}^{(n-1)}(t') - \frac{k}{\hbar} \sum_{n_2} \gamma_2^2 T_{m_2,n_2}(t') e^{j\omega_{m_2,n_2} t'} C_{n_2}^{(n-1)}(t') \right. \\ \left. - \frac{k}{\hbar} \sum_{n_3} \gamma_3^2 T_{m_3,n_3}(t') e^{k\omega_{m_3,n_3} t'} C_{n_3}^{(n-1)}(t') \right] \end{pmatrix} \quad (17)$$

The above equation should be solved in an iterative manner, with the superscript $n$ representing the order of the iteration. If we substitute the first order expressions for the coefficients into the second order ones, a very large (and cumbersome) expression comprising nine terms will result.

*2.5. Transition probabilities and transition rates*

The transition probability ($P(t)_{m_1 \to l_1} = |\gamma_1 C_{m_1}^{(2)}(t)|^2$) and the transition rate of the overall process, which is the time derivative of the transition probability (i.e., $R(t)_{m_1 \to l_1} = d|\gamma_1 C_{m_1}^{(2)}(t)|^2 / dt$), can be determined once the modulus of $\gamma_1 C_{m_1}^{(2)}(t)$ is taken and the appropriate initial conditions are known.

Let´s consider that the neutral, cationic and anionic forms of the molecule are initially in their respective $n_1$, $n_2$ and $n_3$ of states, so that $\left|C_{n_1}^{(0)}(t=0)\right| = \left|C_{n_2}^{(0)}(t=0)\right| = \left|C_{n_3}^{(0)}(t=0)\right| = 1$. While the first order coefficient would be given by

$$
\begin{pmatrix} \gamma_1 C^{(1)}_{m_1}(t) \\ \\ \gamma_2 C^{(1)}_{m_2}(t) \\ \\ \gamma_3 C^{(1)}_{m_3}(t) \end{pmatrix} =
\begin{pmatrix}
\gamma_1 \int_{t_0}^{t} dt' \left[ \frac{-i}{\hbar} \sum_{n_1} \gamma_1^2 T_{m_1,n_1}(t') e^{i\omega_{m_1,n_1} t'} C^{(0)}_{n_1}(t') - \frac{i}{\hbar} \sum_{n_2} \gamma_2^2 T_{m_2,n_2}(t') e^{j\omega_{m_2,n_2} t'} C^{(0)}_{n_2}(t') - \frac{i}{\hbar} \sum_{n_3} \gamma_3^2 T_{m_3,n_3}(t') e^{k\omega_{m_3,n_3} t'} C^{(0)}_{n_3}(t') \right] \\
\gamma_2 \int_{t_0}^{t} dt' \left[ \frac{-j}{\hbar} \sum_{n_1} \gamma_1^2 T_{m_1,n_1}(t') e^{i\omega_{m_1,n_1} t'} C^{(0)}_{n_1}(t') - \frac{j}{\hbar} \sum_{n_2} \gamma_2^2 T_{m_2,n_2}(t') e^{j\omega_{m_2,n_2} t'} C^{(0)}_{n_2}(t') - \frac{j}{\hbar} \sum_{n_3} \gamma_3^2 T_{m_3,n_3}(t') e^{k\omega_{m_3,n_3} t'} C^{(0)}_{n_3}(t') \right] \\
\gamma_3 \int_{t_0}^{t} dt' \left[ \frac{-k}{\hbar} \sum_{n_1} \gamma_1^2 T_{m_1,n_1}(t') e^{i\omega_{m_1,n_1} t'} C^{(0)}_{n_1}(t') - \frac{k}{\hbar} \sum_{n_2} \gamma_2^2 T_{m_2,n_2}(t') e^{j\omega_{m_2,n_2} t'} C^{(0)}_{n_2}(t') - \frac{k}{\hbar} \sum_{n_3} \gamma_3^2 T_{m_3,n_3}(t') e^{k\omega_{m_3,n_3} t'} C^{(0)}_{n_3}(t') \right]
\end{pmatrix} \quad (18)
$$

the expression for the second order coefficient ( $C^{(2)}_{m_1}(t)$ for example) is

$$
\gamma_1 C^{(2)}_{m_1}(t) = \gamma_1 \int_{t_0}^{t} dt' \left[ \frac{-i}{\hbar} \sum_{n_1} \gamma_1^2 T_{m_1,n_1}(t') e^{i\omega_{m_1,n_1} t'} C^{(1)}_{n_1}(t') - \frac{i}{\hbar} \sum_{n_2} \gamma_2^2 T_{m_2,n_2}(t') e^{j\omega_{m_2,n_2} t'} C^{(1)}_{n_2}(t') + - \frac{i}{\hbar} \sum_{n_3} \gamma_3^2 T_{m_3,n_3}(t') e^{k\omega_{m_3,n_3} t'} C^{(1)}_{n_3}(t') \right] \quad (19)
$$

which can be rewritten as

$$
\begin{aligned}
\gamma_1 C^{(2)}_{m_1}(t) = {} & \\
\gamma_1 \int_{t_0}^{t} dt' \int_{t_0}^{t'} dt'' \Bigg[ & \frac{-i}{\hbar} \gamma_1^2 \sum_{l_1} T_{m_1,l_1}(t') e^{i\omega_{m_1,l_1} t'} \left( \frac{-i}{\hbar} \right) \sum_{x=1}^{3} \sum_{n_x} \gamma_x^2 T_{l_x,n_x}(t'') e^{q_x \omega_{l_x,n_x} t''} C^{(0)}_{n_x}(t'') + \\
& \frac{-i}{\hbar} \gamma_2^2 \sum_{l_2} T_{m_2,l_2}(t') e^{i\omega_{m_2,l_2} t'} \left( \frac{-j}{\hbar} \right) \sum_{x=1}^{3} \sum_{n_x} \gamma_x^2 T_{l_x,n_x}(t'') e^{q_x \omega_{l_x,n_x} t''} C^{(0)}_{n_x}(t'') + \\
& \frac{-i}{\hbar} \gamma_3^2 \sum_{l_3} T_{m_3,l_3}(t') e^{i\omega_{m_3,l_3} t'} \left( \frac{-k}{\hbar} \right) \sum_{x=1}^{3} \sum_{n_x} \gamma_x^2 T_{l_x,n_x}(t'') e^{q_x \omega_{l_x,n_x} t''} C^{(0)}_{n_x}(t'') \Bigg]
\end{aligned} \quad (20)
$$

after substituting the first order terms.

*2.6. Adiabatic approximation*

If the coupling perturbation is switched-on adiabatically, we can mathematically consider the limit $t_0 \to -\infty$ (i.e., we admit that no perturbation existed in the distant past) and assume that $T_{m_x,n_x}(t) = T_{m_x,n_x}.e^{\eta.t}$, where $\eta$ is a small positive number to be taken equal to zero at the end of the calculation. For briefness we will work out explicitly only the first three terms to appear in the expression for $C^{(2)}_{m_1}(t)$. (Even though at the end the final result will be shown after all relevant contributions have been taken into account. If one takes the modulus of the nine terms, 81 new ones will result, 72 of them crossing terms of oscillatory behavior from which one should expect no relevant global contribution – see supporting information. Hence, all relevant physical information is in fact contained in the nine remaining terms, the last six of them quite similar to the first three ones, as shown here.) In this manner we have

$$\begin{aligned}
\gamma_1 C^{(2)}_{m_1}(t) = \gamma_1 \int_{-\infty}^{t} dt' \int_{-\infty}^{t'} dt'' \Bigg[ & \frac{\gamma_1^2\gamma_1^2}{\hbar^2} \sum_{n_1}\sum_{l_1} T_{m_1,l_1} e^{i\omega_{m_1,l_1}t'+\eta.t'} T_{l_1,n_1} e^{i\omega_{l_1,n_1}t''+\eta.t''} C^{(0)}_{n_1}(t'') \\
& + \frac{\gamma_1^2\gamma_2^2}{\hbar^2} \sum_{n_2}\sum_{l_1} T_{m_1,l_1} e^{i\omega_{m_1,l_1}t'+\eta.t'} T_{l_2,n_2} e^{j\omega_{l_2,n_2}t''+\eta.t''} C^{(0)}_{n_2}(t'') + \\
& + \frac{\gamma_1^2\gamma_3^2}{\hbar^2} \sum_{n_3}\sum_{l_1} T_{m_1,l_1} e^{i\omega_{m_1,l_1}t'+\eta.t'} T_{l_3,n_3} e^{k\omega_{l_3,n_3}t''+\eta.t''} C^{(0)}_{n_3}(t'') + \ldots \Bigg]
\end{aligned} \tag{21}$$

and, therefore, the transition rate can be written as

$$\begin{aligned}
\gamma_1^2 \frac{d\left|C^{(2)}_{m_1}(t)\right|^2}{dt} \quad & \gamma_1^2 \frac{1}{\hbar^4}\left(\gamma_1^2\gamma_1^2\right)^2 \sum_{n_1}\sum_{l_1} \frac{\left|T_{m_1,l_1}\right|^2 \left|T_{l_1,n_1}\right|^2 4\eta.e^{4\eta.t}}{\left((\omega_{m_1,l_1}+\omega_{l_1,n_1})^2+4\eta^2\right)\left(\omega^2_{l_1,n_1}+\eta^2\right)} + \\
& + \gamma_1^2 \frac{1}{\hbar^4}\left(\gamma_1^2\gamma_2^2\right)^2 \sum_{n_2}\sum_{l_1} \frac{\left|T_{m_1,l_1}\right|^2 \left|T_{l_2,n_2}\right|^2 4\eta.e^{4\eta.t}}{\left(\omega^2_{m_1,l_1}+\omega^2_{l_2,n_2}+4\eta^2\right)\left(\omega^2_{l_2,n_2}+\eta^2\right)} + \\
& + \gamma_1^2 \frac{1}{\hbar^4}\left(\gamma_1^2\gamma_3^2\right)^2 \sum_{n_3}\sum_{l_1} \frac{\left|T_{m_1,l_1}\right|^2 \left|T_{l_3,n_3}\right|^2 4\eta.e^{4\eta.t}}{\left(\omega^2_{m_1,l_1}+\omega^2_{l_3,n_3}+4\eta^2\right)\left(\omega^2_{l_3,n_3}+\eta^2\right)} + \ldots
\end{aligned} \tag{22}$$

*2.7. Separation of terms in different powers of t*

Let's now consider the limit $\eta \to 0$. The usual way to do this [25] is first assume that $e^{4\eta.t} \approx 1$ and only after this to take the desired limit. Here, however, we could alternatively go one step further [26] and take into account the next term in the exponential expansion, i.e., $e^{4\eta.t} \approx 1 + 4\eta.t$, in such manner as to obtain

$$\gamma_1^2 \frac{d\left|C_{m_1}^{(2)}(t)\right|^2}{dt} = \gamma_1^2 \frac{1}{\hbar^4}\left(\gamma_1^2\gamma_1^2\right)^2 \sum_{n_1}\sum_{l_1} \frac{\left|T_{m_1,l_1}\right|^2 \left|T_{l_1,n_1}\right|^2 4\eta.(1+4\eta.t)}{\left((\omega_{m_1,l_1}+\omega_{l_1,n_1})^2+4\eta^2\right)\left(\omega_{l_1,n_1}^2+\eta^2\right)} +$$
$$+\gamma_1^2 \frac{1}{\hbar^4}\left(\gamma_1^2\gamma_2^2\right)^2 \sum_{n_2}\sum_{l_1} \frac{\left|T_{m_1,l_1}\right|^2 \left|T_{l_2,n_2}\right|^2 4\eta.(1+4\eta.t)}{\left(\omega_{m_1,l_1}^2+\omega_{l_2,n_2}^2+4\eta^2\right)\left(\omega_{l_2,n_2}^2+\eta^2\right)} + \qquad (23)$$
$$+\gamma_1^2 \frac{1}{\hbar^4}\left(\gamma_1^2\gamma_3^2\right)^2 \sum_{n_3}\sum_{l_1} \frac{\left|T_{m_1,l_1}\right|^2 \left|T_{l_3,n_3}\right|^2 4\eta.(1+4\eta.t)}{\left(\omega_{m_1,l_1}^2+\omega_{l_3,n_3}^2+4\eta^2\right)\left(\omega_{l_3,n_3}^2+\eta^2\right)} + \ldots$$

before determining the limiting value. After taking the appropriate limit and considering the definition of the Dirac delta function, we can write the final expression as $d\left|C_{m_1}^{(2)}(t)\right|^2 / dt = O_{(0)} + O_{(t)}$, i.e., a sum of a term $O_{(0)}$ that collects all time independent contributions plus a term $O_{(t)}$ that reunites the contributions linear in $t$. In this manner, we have

$$O_{(0)} = \gamma_1^2 \frac{\pi}{\hbar}\left[\left(\gamma_1^2\gamma_1^2\right)^2 \sum_{l_1}\sum_{n_1} \frac{\left|T_{m_1,l_1}\right|^2 \left|T_{l_1,n_1}\right|^2 \delta(E_{m_1}-E_{n_1})}{\left(E_{l_1}-E_{n_1}\right)^2} + \right. \qquad (24)$$
$$+\left(\gamma_1^2\gamma_2^2\right)^2 \sum_{l_1}\left|T_{m_1,l_1}\right|^2 \left(\sum_{n_2=l_2} \frac{\left|T_{l_2,n_2}\right|^2 4\delta(E_{l_2}-E_{n_2})}{\left(E_{l_2}-E_{n_2}\right)^2+\left(E_{m_1}-E_{l_1}\right)^2}\right) +$$
$$\left. +\left(\gamma_1^2\gamma_3^2\right)^2 \sum_{l_1}\left|T_{m_1,l_1}\right|^2 \left(\sum_{n_3=l_3} \frac{\left|T_{l_3,n_3}\right|^2 4\delta(E_{l_3}-E_{n_3})}{\left(E_{l_3}-E_{n_3}\right)^2+\left(E_{m_1}-E_{l_1}\right)^2}\right) + \ldots \right]$$

and

$$O_{(t)} = \gamma_1^2 \frac{8\pi^2 t}{\hbar^2}\left[\left(\gamma_1^2\gamma_1^2\right)^2 \sum_{n_1}\sum_{l_1} \left|T_{m_1,l_1}\right|^2 \left|T_{l_1,n_1}\right|^2 \delta(E_{m_1}-E_{n_1}).\delta(E_{l_1}-E_{n_1}) + \right.$$
$$+\left(\gamma_1^2\gamma_2^2\right)^2 \sum_{n_2}\sum_{l_1} \left|T_{m_1,l_1}\right|^2 \left|T_{l_2,n_2}\right|^2 \delta(E_{m_1}-E_{l_1}+E_{l_2}-E_{n_2}).\delta(E_{l_2}-E_{n_2}) + \qquad (25)$$
$$\left. +\left(\gamma_1^2\gamma_3^2\right)^2 \sum_{n_3}\sum_{l_1} \left|T_{m_1,l_1}\right|^2 \left|T_{l_3,n_3}\right|^2 \delta(E_{m_1}-E_{l_1}+E_{l_3}-E_{n_3}).\delta(E_{l_3}-E_{n_3}) + \ldots \right].$$

Let's interpret carefully each one of the two above contributions. With regard to Eq. (24), its first term corresponds to the usual second order Fermi Golden rule [25], for which transitions involving intermediate states $l_1$ are allowed, provided that the final and initial molecular states are the same. Note that in this case, although the energy is conserved in the entire process, in the intermediary step we must have $E_{l_1} \neq E_{m_1} = E_{n_1}$ (see more on this below). Also, the charge state does not change during the process and the labels remain unchanged (i.e. $m_x$, $n_x$ and $l_x$, all of them with $x=1$, are kept fixed). The novel contributions are the existence of the second and the third terms; each one of them individually mixes the charge state labeled as 1 with each one of the two others (respectively labeled as 2 and 3) and hence they allow the mixture of the different quaternionic sub-spaces. In physical terms, these processes involve a transition from a (initially pure, due to our assumption about the system state before any perturbation is switched on) charge state labeled as 1 to a new charge state (2 or 3). Now the initial and final energies are not the same, and therefore – as it should be expected – there is no conservation of the total energy due to the change in the overall charge state of the molecular system. Note, however, that the condition $\Delta E_1 = (E_{m_1} - E_{l_1}) \neq 0$ is required in these two terms (second and third ones). In this manner, each one of these three contributions requires the occurrence of intermediary transitions to and from molecular states of different energies (and charge states, for the last two terms). Hence, the $O_{(0)}$ terms of Eq. (24) correspond to non-resonant transitions in a particular sub-space (i.e., non-resonant with regard to phonon-induced transitions, for example) and should not be considered in the present description of the molecular conductance problem.

The physical interpretation of the contributions forming the term $O_{(t)}$ (Eq. (25)) is quite similar. The first contribution represents an additional term in the second order Fermi Golden rule [26] that, as discussed above, requires conservation of the energy in transitions involving intermediate states labeled by $l_1$; however, differently to what it was seen in the $O_{(0)}$ case, here it is also required that the energy of the intermediate states be the same as those of the initial and final states ($E_{m_1} = E_{l_1}$ and $E_{l_1} = E_{n_1}$). This is a very interesting term in the context of transport phenomena, because (as we will shortly see) it represents the ballistic (coherent) contribution to the global transmission function. The second and the third terms of $O_{(t)}$, however, are of also of novel character and must be carefully analyzed. As in the $O_{(0)}$ case, the two extra terms allow for a transition from a particular charge state (labeled as 1, due to our initial assumption about the system) to a different charge state (2 or 3). The initial and final energies are not the same, in another indication that the total energy of the system is not conserved since there was a change in its charge state. Although energy conservation on the entire process does not occur, each intermediate process involved corresponds to a resonant (coherent) transition.

This represent the physical situation described as co-tunneling by Ratner and collaborators [15]; once one electron enters the system – in this case, by a resonant transition (i.e., one that involves a close energy matching of the electrode and molecular levels), the molecule automatically changes its charge state and rearranges itself in such a manner that a second electron can leave the system (see bottom panel in Scheme A). This energy rearrangement is sometimes referred to as expressing the need to involve virtual states [15] that are relevant only during the charge transfer process.

Summing over the $m_1$ states, multiplying by the fundamental charge $q$ and taking into account the appropriate Fermi distributions for the occupied and unoccupied states, we can write a generic expression for the electrical current that traverses the molecular system where the ballistic and co-tunneling contributions are simultaneously considered. Hence, in this formalism the relative contributions of the entirely coherent (i.e., only involving the neutral species) and non-coherent (involving the participation of a charged molecular species) terms can be estimated for any chosen value of the applied external electric field. However, before proceeding any further, let's estimate the relative importance of the $O_{(0)}$ and $O_{(t)}$ terms for the molecular conductance problem.

*2.8. Short-time and longer time events*

In terms of the Fermi golden rule, a well-known result [27] is that for extremely short times the probability of a transition is not proportional to $t$ but rather to $t^2$ (and, consequently, its derivative is not independent of $t$). This means that if terms of order $O_{(t)}$ are supposed to occur in extremely short times, then in a given finite time interval $\Delta t$ it can be expected that these terms appear more frequently than those of order $O_{(0)}$. Hence, the latter contribution can be considered associated to less frequent events and the more frequent ones to those of order $O_{(t)}$. This would be an additional reason why, in obtaining an expression for the total electrical current traversing the molecular system, we can neglect the $O_{(0)}$ terms.

Hence, we can now rewrite Eq. (19) retaining only terms involving $O_{(t)}$ as

$$\gamma_1^2 \frac{d\left|C_{m_1}^{(2)}(t)\right|^2}{dt} = \gamma_1^2 \frac{8\pi^2 t}{\hbar^2}\left[\left(\gamma_1^4\right)^2 \sum_{n_1}\sum_{l_1}\left|T_{m_1,l_1}\right|^2\left|T_{l_1,n_1}\right|^2 \delta(E_{m_1}-E_{n_1}).\delta(E_{l_1}-E_{n_1}) + \right.$$
$$+\left(\gamma_1^2\gamma_2^2\right)^2 \sum_{n_2}\sum_{l_1}\left|T_{m_1,l_1}\right|^2\left|T_{l_2,n_2}\right|^2 \delta(E_{m_1}-E_{l_1}+E_{l_2}-E_{n_2}).\delta(E_{l_2}-E_{n_2}) + \qquad (26)$$
$$\left. +\left(\gamma_1^2\gamma_3^2\right)^2 \sum_{n_3}\sum_{l_1}\left|T_{m_1,l_1}\right|^2\left|T_{l_3,n_3}\right|^2 \delta(E_{m_1}-E_{l_1}+E_{l_3}-E_{n_3}).\delta(E_{l_3}-E_{n_3}) + \ldots\right] \quad .$$

Next, we multiply Eq. (26) by the fundamental charge $q$, and insert conveniently the well-known relations [28] for the creation ($a^{\dagger}_{\sigma,n_x}$) and annihilation ($a_{\sigma,n_x}$) Fermion operators acting on the $n^{th}$ level of the $x$-charge state, i.e.: $\left\{a_{\sigma,n_x}, a^{\dagger}_{\lambda,m_x}\right\} = \delta_{m_x,n_x}.\delta_{\sigma,\lambda}$, $\left\langle a^{\dagger}_{\sigma,n_x} a_{\sigma,n_x}\right\rangle = f_{\sigma(E_{n_x})}$ and $\left\langle a_{\sigma,n_x} a^{\dagger}_{\sigma,n_x}\right\rangle = A_{\sigma(E_{n_x})}$, where the subscript σ represents the fragment of the molecular device considered, i.e., the left [right] electrode $(\sigma,\lambda = LE[RE])$ or the extended-molecule itself $(\sigma,\lambda = EM)$. Also, note that we consider each quaternionic sub-space as frozen, in the sense that creation (or destruction) of a given electron is automatically associated to a change in the charge state of the molecule.

Finally, to obtain an expression for the current traversing the "extended molecule" we must convert the summations over the energy levels ($m_x$, $n_x$ with $x = 1, 2, 3$) of the (semi-infinite) electrodes into integrals, since they run over a continuum spectrum, i.e., $\sum \rightarrow \int dE_y.D_y(E_y)$, with $D_y(E_y)$ being the density of states of the $y$-electrode ($y = LE, RE$).

Summing over all intermediary states $l_x$ and averaging in time in the form $(\bar{t})^{-1}\int_0^{\bar{t}} I(t)dt$, we finally obtain an expression for the forward (i.e., from left to right electrode) electrical current as

$$I(\bar{t})^{L\to R} = \tag{27}$$

$$= \frac{4\bar{t}\pi^2 q}{\hbar^2} \cdot \int_{-\infty}^{\infty}\int_{-\infty}^{\infty} dE_{LE}\, dE_{RE} \left(D_{LE}(E_{LE}) D_{RE}(E_{RE}) f_{LE}(E_{LE}) A_{RE}(E_{RE})\right) G_{1x}(E_{LE}, E_{l_1}, E_{l_x}, E_{RE})$$

where

$$G_{1x}(E_{LE}, E_{l_1}, E_{l_x}, E_{RE}) = \gamma_1^2 \sum_{x=1}^{3} \sum_{l_1} \sum_{l_x} \left|T_{\mu_{RE},l_1}\right|^2 \left|T_{l_x,\nu_{LE}}\right|^2 \delta(E_{RE} - E_{l_1} + E_{l_x} - E_{LE}) . \tag{28}$$

$$\cdot \left[ \delta(E_{l_x} - E_{LE}) A_{EM}(E_{l_1}) \left(\gamma_1^2 \gamma_x^2\right)^2 \left[\delta_{l_1,l_x} \delta_{1,x} + (1-\delta_{1,x}) f_{EM}(E_{l_x})\right] \right].$$

In the above expressions, $\delta_{x,y}$ is the Kronecker delta. Naturally, similar expressions could be obtained for $G_{2,x} \equiv G_{2,x}(E_{LE}, E_{l_2}, E_{l_x}, E_{RE})$ and $G_{3,x} \equiv G_{3,x}(E_{LE}, E_{l_3}, E_{l_x}, E_{RE})$. By adopting a quantum chemical calculation (at the density functional level) one can determine all eigenvalues and eigenvectors relevant to the above expression, not only for the neutral species but also to the single-charged cation and single-charged anion, always in the presence of the applied external voltage. For the left and right semi-infinite electrodes standard methods available in solid state physics can be used [6].

Once a perturbation is introduced in a previously isolated system, off diagonal terms involving different charge states may become relevant. One can then recur to the above expressions to calculate the electrical current flowing through the connected molecule as a function of the applied external bias and, after taking the numerical derivative of this curve, determine the corresponding molecular conductance. Appropriate examples for the cases of conjugated and non-conjugated will be discussed in Refs. [16] and[17], respectively.

Note, however, that a transition between two different charge states must involve change in the energy from the initial to the final state. Hence, following the procedure adopted by Matveev and co-workers [29], we first define an occupation number and then a 'molecular Fermi level' for each one of the three charge states of the extended molecule *EM*. In practice, we must just take into account the new electronic distribution of the transient molecular species formed and define the "molecular Fermi level" characteristic of a charge state $x$ as being halfway between the energies of its Highest Occupied Molecular Orbital [Lowest Unoccupied Molecular Orbital] (HOMO [LUMO]), i.e.,

$$\mu^{x}_{mol} = \frac{E^x_{HOMO} + E^x_{LUMO}}{2} , \tag{29}$$

where $E^x_{HOMO}\left[E^x_{LUMO}\right]$ is the energy of the HOMO [LUMO] of the "$x$-charged extended molecule" in presence of the external bias. It is important to stress that, in doing this, we are assuming the presence of a "contact resistance" [4], so that the Fermi level at each infinite electrode, far from the molecule itself, varies linearly (and are displaced in opposite ways) with the strength of the applied electric field; however, the "molecular Fermi level" as defined above can be adjusted accordingly to a full molecular *ab initio* calculation repeated for each value of electric field [6, 30]. This procedure assures that typical quantum chemical effects are properly described, such as the occurrence of "avoided crossing" situations between two neighboring molecular levels [31]. As we will discuss in forthcoming papers dealing with the application of the present formalism to different systems, the inclusion of such effects provides an entirely new standpoint to interpret phenomena as the occurrence of negative differential resistance (NDR) and the appearance of charge blockade caused by the presence of spatially localized molecular orbitals instead of the canonical Coulomb blockade [16, 17].

*2.9. Taking into Account the Remaining Terms*

Although it is not straightforward to demonstrate, after obtaining the additional terms ($G_{2,x}$ and $G_{3,x}$) of Eq. (28) and summing the results of all sub-spaces (and considering all off-diagonal terms in the derivative $d\left(\gamma_1^2\left|C_{m_1}^{(2)}(t)\right|^2+\gamma_2^2\left|C_{m_2}^{(2)}(t)\right|^2+\gamma_3^2\left|C_{m_3}^{(2)}(t)\right|^2\right)/dt$), the expression for the total current in the device will take the form

$$I(\bar{t})^{L\to R} = \\ = \frac{4\bar{t}\pi^2 q}{\hbar^2}.\int_{-\infty}^{\infty}\int_{-\infty}^{\infty} dE_{LE}dE_{RE}\left(D_{LE}(E_{LE})D_{RE}(E_{RE})f_{LE}(E_{LE})A_{RE}(E_{RE})\right)G(E_{LE},E_{l_y},E_{l_x},E_{RE}), \tag{30a}$$

where

$$G(E_{LE},E_{l_y},E_{l_x},E_{RE}) = \sum_{x,y=1}^{3}\sum_{l_x}\sum_{l_y}\delta(E_{RE}-E_{l_y}+E_{l_x}-E_{LE}).\left|T_{\mu_{RE},l_y}\right|^2\left|T_{l_x,\nu_{LE}}\right|^2\cdot \\ .\ \delta(E_{l_x}-E_{LE})A_{EM}(E_{l_y})\left(\gamma_x^2\gamma_y^2\right)^2\left[\delta_{l_y,l_x}\delta_{y,x}+(1-\delta_{y,x})f_{EM}(E_{l_x})\right]\ . \tag{30b}$$

Let´s then discuss the physical meaning of each individual term in the above expression. To do this we have to analyze both the $x = y$ and $x \neq y$ cases. When $x = y$, the molecule remains in the same charge state during the electron transfer from the left to the right electrode and the corresponding expression represents a transport process through an unoccupied sate of the system considered (and the Pauli exclusion principle is therefore dutifully respected), i.e., a "hot electron", originated deep inside the cathode, crosses the entire extension of the molecule without scattering before thermalizing inside the anode. This is the condition for a coherent (or ballistic) transport as the presence of the two Dirac delta functions in Eqs. (30) (with $x = y$) assures that the total energy is conserved. One could say that in this case the molecule does not "feel" the presence of an extra electron, since the role of the unoccupied states is limited to serve as electron transport channels and the molecular system acts only as a passive element in the device.

On the other hand, cases where $x \neq y$ correspond to a change in the charge state of the molecule. One can associate these instances to the processes dubbed as co-tunneling, which involve situations of a net charge transport between the metallic electrodes, while no coherence between the tunneling events occurring at each of the two junctions is assumed [11]. In case of an effective electron transport between the electrodes, two different electrons are involved: under nonzero bias, while one electron tunnels from the cathode into the system – which is initially in a charge state labeled by *x* (let's say, neutral) – another one is transferred from the system – now in a modified charge state labeled by *y* (anion, in this example) – to the anode. In this manner, during a co-tunneling event the extended molecule changes its charge state and assumes an "active behavior", with its electronics properties (energy and wavefunction) adjusting themselves to the electron flow between the electrodes.

Obviously the process described here defines the *electronic* current in a specific direction, i.e., from left to right, for example, or $I(\bar{t})^{L\rightarrow R}$. Naturally, the net current is the difference $I_{tot} = I(\bar{t})^{L\rightarrow R} - I(\bar{t})^{R\rightarrow L}$ and the conductance could be obtained by taking the numerical derivative of the above expression. Once again, we stress the fact that both ballistic and co-tunneling processes are allowed to contribute to the overall charge transport in a given system, and so the above formalism is uniquely suited to deal with the full range of coupling regimes, i.e., it can be applied to weakly, intermediate and strongly coupled systems.

### *2.10. Time-scale of the ballistic and co-tunneling processes*

An important aspect of the electron transport process to be considered concerns the typical time scales involved: even though the co-tunneling and ballistic transport are very different in nature, both take similar amounts of time to occur. This seemly paradoxical statement can be better understood once one realizes that in the ballistic case a given electron must "cross" the entire system, while in a co-tunneling event – although the system must relax to a different charge state once a first electron is transferred – an electron already on the "other side" of the molecule can immediately tunnel to the second electrode. Then, at least in principle one could estimate the "relaxation time" of the new charge state either by calculating the time spent for a ballistic electron to cross the system or by using the uncertainty principle after considering the finite lifetime of the transient ionic state [32].

### *2.11. Relative probabilities of the ballistic and co-tunneling processes*

So far, we have seen that during the electron transport between the metallic electrodes, the intersected molecular system may (in case of co-tunneling/non-coherent events) or may not (for ballistic/coherent ones) change its charge state. So the question arises of how to estimate the probability for these two physical situations to occur?

In fact, we still have a free choice for the parameter $\gamma_x$ (x = 1, 2, 3) that was only briefly mentioned before. Note that only the square of this parameter is involved in the above expressions, so that it can be taken as the phenomenological probability of finding the system in a particular charge state. To estimate the possible range of values for this parameter, we remember that a necessary condition is that the quaternionic (multi charge) state (see Eq. (4)) be normalized ($\gamma_1^2 + \gamma_2^2 + \gamma_3^2 = 1$). Also we expect that the chemical structure of the molecule must play an important role in the process whatever the kind of dominant transport mechanism. In other words, as confirmed by the known experimental results, the chemical structure of the molecule will determine the relative contributions of the coherent and non-coherent terms to the overall charge transport. It is well known that while conjugated organic molecules exhibit very spatially delocalized π-orbitals, the localized molecular orbitals associated to the presence of a saturated group in a bridge position connecting donor and acceptor moieties in a given molecule can act as an effective barrier hindering the electron transfer from one side of the non-conjugated molecule to the other. Then, as a natural choice, we will associate the facility of a molecular system in allowing charge delocalization to its degree of *aromaticity* [33], and set the value of the parameter $\gamma_x$ as an alternative manner of estimating the aromaticity of an organic structure. One could expect that in more aromatic molecules the ballistic term will give the largest relative contribution to the overall electrical transport; by a similar reasoning, co-tunneling terms must dominate the electron transfer between terminal electrodes that connect molecules with a saturated structure.

Assuming that the trapping of a charge within a molecular system during a period of time that allows for the rearrangement of its internal structure will be a rare (and discrete) event, we will associate the term $\gamma_x$ to a Poisson distribution [34], i.e., the probability $\gamma_x^2$ will have the form

$$\gamma_x^2 = \frac{e^{-\varsigma}}{k!}(\varsigma)^k \qquad \text{where} \qquad \varsigma = \frac{1}{N}\sum_{i=1}^{N}\left(\frac{R_i}{R_m}-1\right)^2 \qquad (31)$$

Note that in the above equation: *i)* $R_m$=1.39739 $\mathring{A}$ was adopted as the calculated average bond length for a benzene molecule terminally substituted with thiol groups at opposite ends as determined by an optimization calculation at the DFT level using the B3LYP functional and a 6-31G(d,p) basis set [35], i.e., the result of the division of the sum of all carbon-carbon bond lengths by the number of bonds in the molecule, *ii)* $k$ is the number of electrons (or holes) "trapped" by the molecule (and therefore, while $k=0$ for the case of a neutral molecule, $k=1$ for a single charged anion or cation), and *iii)* the summing over $k$ allows the normalization of the Poisson distribution. Hence, for this *anzats* to be valid when only three charge states are relevant to the problem, the sum over these states must be almost 1, i.e., we must have $\gamma_1^2 + \gamma_2^2 + \gamma_3^2 < a \approx 1$ . (To go beyond this approximation and include more charged states, we must consider octanions [19], for example.) With the above definitions, if we choose $x = 2$ to represent anions, then $k = 1$ would correspond to a single-charged molecular anion, while $k = 2$ would represent the doubly charged anion. Naturally, when $x = 1$ (neutral species) we must have $k = 0$, so that the product $\gamma_1^2.\gamma_1^2.\gamma_1^2.\gamma_1^2$ represents a ballistic event in which the extended molecule remains neutral.

An alternative manner of interpreting the *anzats* of adopting a Poisson distribution for the probabilities $\gamma_x^2$ is to assume a phenomenological point of view: for example, to take into account the fact that after an electron leaves the left electrode it has the possibility to move to the molecule in three different charge states, the hopping to each particular charge state will be described by a probability defined by the parameter $\varsigma$ . For the neutral case, the parameter $k$ will be zero, while for both cationic and anionic species, $k = 1$. This is equivalent to say that we are not able to distinguish whether the molecule captures an electron or a hole: all internal processes that could occur inside the molecule are not accessible to us and will be described in a phenomenological way by the Poisson probability.

As a final note, we observe that the quaternionic time dependent perturbation treatment here developed could be viewed as a general formalism that may find applications to different problems where one needs to determine the transitions rate of any perturbed system that can evolve into a three-pronged manifold. Coupled to time-dependent perturbation techniques, the quaternions formalism (and its non-commutative algebra) can be used as a powerful mathematical tool whenever up to two additional physically related possibilities exist for the time evolution of a physical system. Examples in quantum chemistry problems would include, among others, the description of tautomers and *cis-trans* isomers, and the analyses of protonation or oxidation-reduction processes.

## 3. Conclusion

In this work we presented a new theoretical approach to calculate the electrical current that flows through a molecule connected to two terminal electrodes and subject to an externally applied electric field. We allow for the possibility of both ballistic and non-coherent transport, and we call attention to the active role that the molecular levels play especially in the latter case, where the molecular conductance profile is particularly sensitive to the opening and closing of transport channels through the transient single-charged species. For this, we have chosen to use time dependent perturbation theory within a quaternion formalism, where the time evolution of an initially neutral molecule can proceed either preserving the charge or by following the Hamiltonians that describe a single-charged anion or a single-charged cation. In this manner, we go beyond canonical time dependent formalism since the three quaternionic subspaces (each one with its own imaginary unit and corresponding to the three different charge states of the molecular system) can now be coupled. (We call attention to the fact that, although the application of the quaternionic formalism to other physical situation is not the focus of the present paper, it is straightforward to extend the treatment to show that the use of a time ordering operator and related Dyson series can be naturally associated to a quaternionic system. In these cases one obtain more general expressions that reduce to the known results for the case where just one quaternionic sub-space is taken into account. The corresponding details are presented in the final appendices of the accompanying support information.)

The main advantage of the present approach is that in the final expression for the electrical current flowing through the molecule the ballistic (coherent) and co-tunneling (incoherent) terms appear naturally as complementary regimes of transport. Also, the treatment can take into account the possibility of describing strongly correlated situations (as those associated to the presence of spatially localized molecular orbitals) where the ballistic regime usually fails [4].

In forthcoming papers we will discuss the application of the present formalism to cases where ballistic [16] and non-coherent [17] processes become important at specific regimes of the externally applied bias, as well as to aminoacid related systems, for which both mechanisms compete [36]. Although this is just a preliminary work, we believe that it introduces a new and general description of the transport phenomena in molecular systems where a non-zero possibility exists for the capture of an electron or a hole from the connected electrodes, changing the charge of the molecule and allowing the process to continue in a new potential surface. This will give rise to new possibilities of approaching and interpreting the molecular conductance problem.

Acknowledgments: This work had the financial support from the Brazilian agencies CNPq and FINEP and from INFo, one of the excellence institutes of the Brazilian Ministry of Science and Technology INCT program. ACLM would like to thank CAPES for a graduate fellowship.

*corresponding author: celso@df.ufpe.br

Scheme A

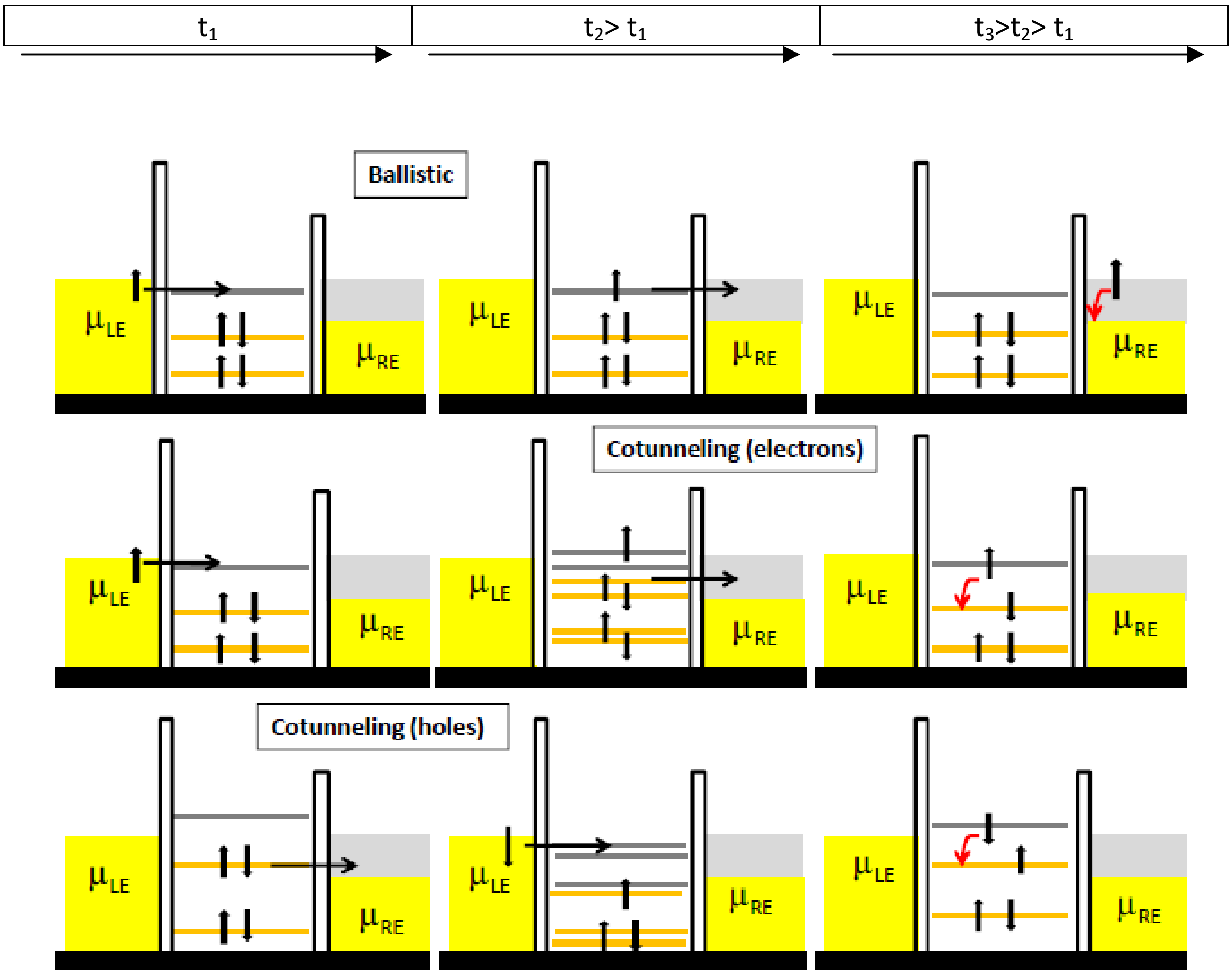

Top panel: ("coherent" or ballistic mode; extended molecule remains neutral - EM)
Middle panel: (co-tunneling of electrons, a non-coherent mode; transient anion – $EM^-$)
Bottom panel: (co-tunneling of holes, a non-coherent mode; transient cation – $EM^+$)

Scheme A: graphical representation of alternative mechanisms of charge transport between the left and right semi-infinite metallic electrodes. Note that in the non-coherent modes the extended molecule plays an active part in the charge flow process.

Figure Caption

Fig. 1 - Schematic representation of the extended molecule $EM \equiv C_L - M - C_R$ concept: a full quantum-chemical *ab initio* calculation is performed for the molecule of interest (M) while plus two connected gold metal clusters ($C_L$ and $C_R$). The EM is then assumed to be in electrical contact with the two semi-infinite metallic electrodes via fixed coupling parameter.

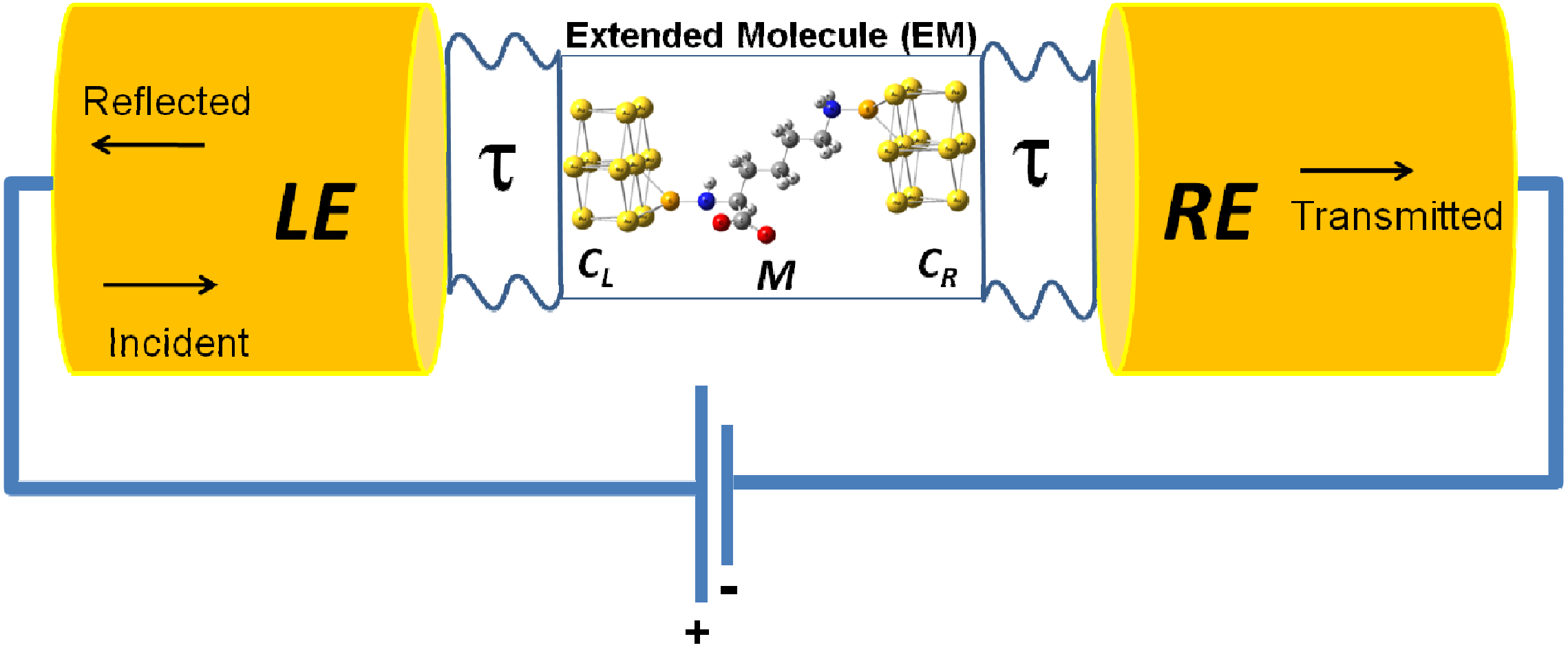

Figure 1

# A Quaternion Based Quantum Chemical *ab initio* Treatment of Coherent and Non-Coherent Electron Transport in Molecules

Augusto C. L. Moreira and Celso P. de Melo*

Departamento de Física, Universidade Federal de Pernambuco

50670-901 Recife, PE, Brazil

## Supporting Information

# Appendix 1: Brief Review of the Quaternions Concept

In this Appendix we will review the main Quaternion results used in this work [1, 2]. The literature concerning quaternions is very extensive and to people who want to learn more about this topic, we recommend examining the material listed in the accompanying bibliography [3].

Strictly speaking, the formalism of Quaternions was devised by William Rowan Hamilton on the 16th of October in 1843. A Quaternion ***q*** could be considered as a 4-component extended complex number belonging to a set $\mathcal{H}$ of the form: $q = q_0 + iq_1 + jq_2 + kq_3$, where $q_0$, $q_1$, $q_2$ and $q_3$ are real numbers and the imaginary components (*i, j, k*) satisfy the following rules

$$i^2 = j^2 = k^2 = ijk = -1$$

$$ij = -ji = k \quad , \qquad jk = -kj = i \quad , \qquad ki = -ik = j \qquad ,$$

of a non-commutative algebra.

The real and imaginary (or vector) parts of $q$ are $\mathrm{Re}(q) = q_0$ and $\mathrm{Im}(q) = iq_1 + jq_2 + kq_3$, respectively. The summation of quaternions follows the same rules as vector sums: thus if we define another quaternion $p = p_0 + ip_1 + jp_2 + kp_3$, the $p + q$ sum is given by $p + q = (p_0 + q_0) + i(p_1 + q_1) + j(p_2 + q_2) + k(p_3 + q_3)$.

In analogy with the algebra of complex numbers, we can define the conjugate of a quaternion as $\bar{q} = q_0 - iq_1 - jq_2 - kq_3$, so that for any pair of quaternion numbers *p, q*, the conjugate of the product is given by $\overline{qp} = \bar{p}\bar{q}$, i.e., the conjugate of a product is the product of the conjugates in reverse order.

The inner product of two quaternions is $\langle p,q\rangle = p\bar{q} = p_0q_0 + p_1q_1 + p_2q_2 + p_3q_3$. With this definition, the quaternion set $\mathcal{H}$ can be considered to define a 4-dimensional real Hilbert space. We can define the norm of a quaternion as $N(q) = |q| = \sqrt{|q|^2} = \sqrt{q\bar{q}} = \sqrt{q_0^2 + q_1^2 + q_2^2 + q_3^2}$ (note that $\bar{q}q = q\bar{q} = |q|^2$), and a quaternion is called unitary if its norm is $N(q) = |q| = 1$.

If $q \neq 0$ we can define the inverse of quaternion as $q^{-1} = \bar{q}/|q|^2$, so that $q^{-1}q = qq^{-1} = 1$ . Thus, with the sum and product operations as defined above, the set $\mathcal{H}$ forms a division ring.

Other ways of representing quaternions are:

i) Vector form: $q = q_0 + \vec{h}.\vec{q}$, where $\vec{h} = (i, j, k)$ and $\vec{q} = (q_1, q_2, q_3)$ with the dot denoting the usual scalar product. Note that, in the vectorial notation we have $pq = p_0q_0 - (\vec{p}.\vec{q}) + p_0\vec{q} + q_0\vec{p} + \vec{p}\times\vec{q}$ where $\times$ denotes the usual vector product.

ii) Polar form: $q = |q|e^{I\theta}$, where $I = \dfrac{(\vec{h}.\vec{q})}{|\vec{q}|}$ and $\tan(\theta) = \dfrac{|\vec{h}.\vec{q}|}{q_0}$ , $(0 \le \theta \le \pi)$.

iii) symplectic form, a very useful one: $q = z + jw$, where $z = q_0 + iq_1$ and $w = q_2 - iq_3$. Note that in this form the four real components of a quaternion are rewritten as two complexes numbers.

Other definitions are:

1) A quaternionic function is a map that relates each $q \in \mathcal{H}$ to a unique number $w = F(q)$ ).

2) Exponential: $e^q = e^{q_0}\left\{\cos\left(|\vec{q}|\right) + \dfrac{\vec{q}}{|\vec{q}|}\sin\left(|\vec{q}|\right)\right\}$ (Note that $|e^q| = e^{q_0}$ ).

3) Logarithm: $\ln(q) = \ln\left(|q|\right) + \dfrac{\vec{q}}{|\vec{q}|}\cos^{-1}\left(\dfrac{q_0}{|q|}\right)$.

Observations: Also note that neither $e^{p+q} = e^p e^q$ nor $\ln(pq) = \ln(p) + \ln(q)$ are necessarily equal. In general we have $e^{p+q} \neq e^p e^q \neq e^q e^p$ and $\ln(pq) \neq \ln(p) + \ln(q)$.

# Appendix 2: Idempotency of $Q_\alpha$.

Let´s show that $(Q_\alpha)^2 = Q_\alpha$. The easiest way to do it is noting that the quaternionic wave function $\left|\chi_\alpha(t)\right\rangle = \begin{pmatrix} \gamma_1 \left|\psi_{n_1}(t)\right\rangle \\ \gamma_2 \left|\psi_{n_2}(t)\right\rangle \\ \gamma_3 \left|\psi_{n_3}(t)\right\rangle \end{pmatrix}$ ,

(see Eq. (2) of the main text) is normalized, i.e., $\left\langle \chi_\alpha(t) \middle| \chi_\alpha(t) \right\rangle = \gamma_1^2 + \gamma_2^2 + \gamma_3^2 = 1$ .

With this property (normalized wave function) in mind one can write:

$$\left(Q_\alpha(t)\right)^2 = \left|\chi_\alpha(t)\right\rangle\left\langle\chi_\alpha(t)\middle|\chi_\alpha(t)\right\rangle\left\langle\chi_\alpha(t)\right| = \left|\chi_\alpha(t)\right\rangle\left\langle\chi_\alpha(t)\right|\left(\gamma_1^2+\gamma_2^2+\gamma_3^2\right) = \left|\chi_\alpha(t)\right\rangle\left\langle\chi_\alpha(t)\right| = Q_\alpha(t)$$

The same result could be obtained by direct multiplication of $Q_\alpha$ in a matrix form, i.e.:

$$Q_\alpha(t) = \begin{pmatrix} \gamma_1 \left|\psi_{n_1}(t)\right\rangle\left\langle\psi_{n_1}(t)\right|\gamma_1 & \gamma_1 \left|\psi_{n_1}(t)\right\rangle\left\langle\psi_{n_2}(t)\right|\gamma_2 & \gamma_1 \left|\psi_{n_1}(t)\right\rangle\left\langle\psi_{n_3}(t)\right|\gamma_3 \\ \gamma_2 \left|\psi_{n_2}(t)\right\rangle\left\langle\psi_{n_1}(t)\right|\gamma_1 & \gamma_2 \left|\psi_{n_2}(t)\right\rangle\left\langle\psi_{n_2}(t)\right|\gamma_2 & \gamma_2 \left|\psi_{n_2(t)}\right\rangle\left\langle\psi_{n_3}(t)\right|\gamma_3 \\ \gamma_3 \left|\psi_{n_3}(t)\right\rangle\left\langle\psi_{n_1}(t)\right|\gamma_1 & \gamma_3 \left|\psi_{n_3}(t)\right\rangle\left\langle\psi_{n_2}(t)\right|\gamma_2 & \gamma_3 \left|\psi_{n_3(t)}\right\rangle\left\langle\psi_{n_3}(t)\right|\gamma_3 \end{pmatrix}$$

$$\left(Q_\alpha(t)\right)^2 = \begin{pmatrix} \left(\gamma_1 \left|\psi_{n_1}(t)\right\rangle\left\langle\psi_{n_1}(t)\right|\gamma_1\right)\left(\sum_{n=1}^{3}\gamma_n^2\right) & \left(\gamma_1 \left|\psi_{n_1}(t)\right\rangle\left\langle\psi_{n_2}(t)\right|\gamma_2\right)\left(\sum_{n=1}^{3}\gamma_n^2\right) & \left(\gamma_1 \left|\psi_{n_1}(t)\right\rangle\left\langle\psi_{n_3}(t)\right|\gamma_3\right)\left(\sum_{n=1}^{3}\gamma_n^2\right) \\ \left(\gamma_2 \left|\psi_{n_2}(t)\right\rangle\left\langle\psi_{n_1}(t)\right|\gamma_1\right)\left(\sum_{n=1}^{3}\gamma_n^2\right) & \left(\gamma_2 \left|\psi_{n_2}(t)\right\rangle\left\langle\psi_{n_2}(t)\right|\gamma_2\right)\left(\sum_{n=1}^{3}\gamma_n^2\right) & \left(\gamma_2 \left|\psi_{n_2(t)}\right\rangle\left\langle\psi_{n_3}(t)\right|\gamma_3\right)\left(\sum_{n=1}^{3}\gamma_n^2\right) \\ \left(\gamma_3 \left|\psi_{n_3}(t)\right\rangle\left\langle\psi_{n_1}(t)\right|\gamma_1\right)\left(\sum_{n=1}^{3}\gamma_n^2\right) & \left(\gamma_3 \left|\psi_{n_3}(t)\right\rangle\left\langle\psi_{n_2}(t)\right|\gamma_2\right)\left(\sum_{n=1}^{3}\gamma_n^2\right) & \left(\gamma_3 \left|\psi_{n_3(t)}\right\rangle\left\langle\psi_{n_3}(t)\right|\gamma_3\right)\left(\sum_{n=1}^{3}\gamma_n^2\right) \end{pmatrix}$$

Since

$$\sum_{n=1}^{3}\gamma_n^2 = 1 \quad ,$$

we can easily see that $(Q_{\alpha(t)})^2 = Q_{\alpha(t)}$.

## Appendix 3: The double integral.

We will make an explicit calculation of the double integral that appears in Eq. (21) of the main text

$$C_{m_1}^{(2)}(t)=\int_{-\infty}^{t}dt'\int_{-\infty}^{t'}dt''\left[\frac{1}{\hbar^2}\gamma_1^2\gamma_1^2\sum_{n_1}\sum_{l_1}T_{m_1,l_1}e^{i\omega_{m_1,l_1}t'+\eta.t'}T_{l_1,n_1}e^{i\omega_{l_1,n_1}t''+\eta.t''}\right.$$

$$+\frac{1}{\hbar^2}\gamma_1^2\gamma_2^2\sum_{n_2}\sum_{l_1}T_{m_1,l_1}e^{i\omega_{m_1,l_1}t'+\eta.t'}T_{l_2,n_2}e^{j\omega_{l_2,n_2}t''+\eta.t''}$$

$$\left.+\frac{1}{\hbar^2}\gamma_1^2\gamma_3^2\sum_{n_3}\sum_{l_1}T_{m_1,l_1}e^{i\omega_{m_1,l_1}t'+\eta.t'}T_{l_3,n_3}e^{k\omega_{l_3,n_3}t''+\eta.t''}+\dots\right].$$

The first term can be considered as canonical in the sense that there is only one imaginary unity so that all terms in this particular double integral commute.

We will now focus our attention on the second and the third double integrals that have two different imaginary units. In fact, we will just solve in details the second one, as the same reasoning can be applied for analyzing the third one. To do this we initially rewrite it as

$$C_{m_1}^{(2)}(t)=\frac{1}{\hbar^2}(\gamma_1\gamma_2)^2\sum_{n_2}\sum_{l_1}\left(T_{m_1,l_1}\right)\left(T_{l_2,n_2}\right)\int_{-\infty}^{t}dt'\int_{-\infty}^{t'}dt''\left(e^{i\omega_{m_1,l_1}t'+\eta.t'}\right)\left(e^{j\omega_{l_2,n_2}t''+\eta.t''}\right)\ .$$

The first integral can be easily calculated, so that

$$I_{12}(t)=\frac{1}{\hbar^2}(\gamma_1\gamma_2)^2\sum_{n_2}\sum_{l_1}\left(T_{m_1,l_1}\right)\left(T_{l_2,n_2}\right)\int_{-\infty}^{t}dt'\left(e^{i\omega_{m_1,l_1}t'+\eta.t'}\right)\left(\frac{e^{j\omega_{l_2,n_2}t'+\eta.t'}}{j\omega_{l_2,n_2}+\eta}\right)\ .$$

Now, if we focus in the second parentheses, one can note that the real exponential term commutes with all complex exponentials so that we can rewrite the remaining imaginary exponential by use of the Euler identity $e^{ja}=\cos(a)+j\sin(a)$ in the form

$$I_{12}(t)=$$

$$=\frac{1}{\hbar^2}(\gamma_1\gamma_2)^2\sum_{n_2}\sum_{l_1}\left(T_{m_1,l_1}\right)\left(T_{l_2,n_2}\right)\int_{-\infty}^{t}dt'\left(e^{i\omega_{m_1,l_1}t'+2\eta.t'}\right)\left(\frac{\cos(\omega_{l_2,n_2}t')+jsin(\omega_{l_2,n_2}t')}{j\omega_{l_2,n_2}+\eta}\right)$$

We will now separate the above integral in two others, one containing the sine and the other containing the cosine term, i.e.,

$$I_{12a}(t)=\frac{1}{\hbar^2}(\gamma_1\gamma_2)^2\sum_{n_2}\sum_{l_1}\left(T_{m_1,l_1}\right)\left(T_{l_2,n_2}\right)\int_{-\infty}^{t}dt'\left(\cos(\omega_{l_2,n_2}t').e^{i\omega_{m_1,l_1}t'+2\eta.t'}\right)\left(\frac{1}{j\omega_{l_2,n_2}+\eta}\right)$$

and

$$I_{12b}(t)=\frac{1}{\hbar^2}(\gamma_1\gamma_2)^2\sum_{n_2}\sum_{l_1}\left(T_{m_1,l_1}\right)\left(T_{l_2,n_2}\right)\int_{-\infty}^{t}dt'\left(\sin(\omega_{l_2,n_2}t').e^{i\omega_{m_1,l_1}t'+2\eta.t'}\right)\left(\frac{j}{j\omega_{l_2,n_2}+\eta}\right).$$

Since the *j* terms do not commute with those containing *i*, we will keep all terms containing the *j* imaginary unit in the far right hand side, and use the well-known relations

$$\int\cos(ax).e^{bx}dx=\frac{\left(b\cos(ax)+a\sin(ax)\right).e^{bx}}{b^2+a^2}\quad\text{and}\quad\int\sin(ax).e^{bx}dx=\frac{\left(b\sin(ax)-a\cos(ax)\right).e^{bx}}{b^2+a^2}$$

,

to rewrite

$$I_{12a}(t)=\beta_{12}e^{i\omega_{m_1,l_1}t+2\eta.t}\left(\frac{\left((i\omega_{m_1,l_1}+2\eta)\cos(\omega_{l_2,n_2}t)+(\omega_{l_2,n_2})\sin(\omega_{l_2,n_2}t)\right)}{(i\omega_{m_1,l_1}+2\eta)^2+(\omega_{l_2,n_2})^2}\right)\left(\frac{1}{j\omega_{l_2,n_2}+\eta}\right)$$

$$I_{12b}(t)=\beta_{12}e^{i\omega_{m_1,l_1}t+2\eta.t}\left(\frac{\left((i\omega_{m_1,l_1}+2\eta)\sin(\omega_{l_2,n_2}t)-(\omega_{l_2,n_2})\cos(\omega_{l_2,n_2}t)\right)}{(i\omega_{m_1,l_1}+2\eta)^2+(\omega_{l_2,n_2})^2}\right)\left(\frac{j}{j\omega_{l_2,n_2}+\eta}\right)$$

where $\beta_{12}=\frac{1}{\hbar^2}(\gamma_1\gamma_2)^2\left(T_{m_1,l_1}\right)\left(T_{l_2,n_2}\right)$ and, for the sake of simplicity, we have excluded the sums for a moment. Note that now we can write the exponential term at the beginning of the expression because only the same imaginary unit *i* is present.

Now, all that is left to be done is to sum the both integral expressions above and, after taking notice of the non-commutative property of quaternions, to rearrange the terms in a more convenient way, in the form

$$I_{12}(t)=I_{12a}(t)+I_{12b}(t)\ \text{ or}$$

$$I_{12}(t)=\beta_{12}e^{i\omega_{m_1,l_1}t+2\eta.t}\left[\frac{\left((i\omega_{m_1,l_1}+2\eta)\cos(\omega_{l_2,n_2}t)+(\omega_{l_2,n_2})\sin(\omega_{l_2,n_2}t)\right)}{(i\omega_{m_1,l_1}+2\eta)^2+(\omega_{l_2,n_2})^2}\right.$$
$$\left.+\frac{\left((i\omega_{m_1,l_1}+2\eta)\sin(\omega_{l_2,n_2}t)-(\omega_{l_2,n_2})\cos(\omega_{l_2,n_2}t)\right)}{(i\omega_{m_1,l_1}+2\eta)^2+(\omega_{l_2,n_2})^2}.j\right]\left(\frac{1}{j\omega_{l_2,n_2}+\eta}\right).$$

Note that although the denominator does not commute with $j$ in the second term inside the parentheses, it can be placed in evidence at the left hand side,

$$I_{12}(t)=\frac{\beta_{12}e^{i\omega_{m_1,l_1}t+2\eta.t}}{(i\omega_{m_1,l_1}+2\eta)^2+(\omega_{l_2,n_2})^2}\Big[(i\omega_{m_1,l_1}+2\eta)\cos(\omega_{l_2,n_2}t)+(\omega_{l_2,n_2})\sin(\omega_{l_2,n_2}t)$$
$$+(i\omega_{m_1,l_1}+2\eta)j\sin(\omega_{l_2,n_2}t)-(\omega_{l_2,n_2})j\cos(\omega_{l_2,n_2}t)\Big]\left(\frac{1}{j\omega_{l_2,n_2}+\eta}\right).$$

Re-arranging terms we get

$$I_{12}(t)=\frac{\beta_{12}e^{i\omega_{m_1,l_1}t+2\eta.t}}{(i\omega_{m_1,l_1}+2\eta)^2+(\omega_{l_2,n_2})^2}\cdot\Big[i\omega_{m_1,l_1}(\cos(\omega_{l_2,n_2}t)+j\sin(\omega_{l_2,n_2}t))+$$
$$+\omega_{l_2,n_2}(\sin(\omega_{l_2,n_2}t)-j\cos(\omega_{l_2,n_2}t))+2\eta(\cos(\omega_{l_2,n_2}t)+j\sin(\omega_{l_2,n_2}t))\Big]\cdot\left(\frac{1}{j\omega_{l_2,n_2}+\eta}\right).$$

Remembering that $1=-(j)^2$, we can rewrite $\sin(\omega_{l_2,n_2}t)=-(j)^2\sin(\omega_{l_2,n_2}t)$, and hence

$$I_{12}(t)=\frac{\beta_{12}e^{i\omega_{m_1,l_1}t+2\eta.t}}{(i\omega_{m_1,l_1}+2\eta)^2+(\omega_{l_2,n_2})^2}\Big[i\omega_{m_1,l_1}(\cos(\omega_{l_2,n_2}t)+j\sin(\omega_{l_2,n_2}t))+$$
$$+\omega_{l_2,n_2}(-(j)^2\sin(\omega_{l_2,n_2}t)-j\cos(\omega_{l_2,n_2}t))+2\eta(\cos(\omega_{l_2,n_2}t)+j\sin(\omega_{l_2,n_2}t))\Big]\left(\frac{1}{j\omega_{l_2,n_2}+\eta}\right).$$

We can now rewrite

$$I_{12}(t) = \frac{\beta_{12} e^{i\omega_{m_1,l_1}t + 2\eta.t}}{(i\omega_{m_1,l_1} + 2\eta)^2 + (\omega_{l_2,n_2})^2}\Big[ i\omega_{m_1,l_1}(\cos(\omega_{l_2,n_2}t) + j\sin(\omega_{l_2,n_2}t)) +$$

$$- j\omega_{l_2,n_2}(\cos(\omega_{l_2,n_2}t) + j\sin(\omega_{l_2,n_2}t)) + 2\eta(\cos(\omega_{l_2,n_2}t) + j\sin(\omega_{l_2,n_2}t))\Big]\left(\frac{1}{j\omega_{l_2,n_2} + \eta}\right)$$

and group all sine and cosine terms to get

$$I_{12}(t) =$$

$$= \frac{\beta_{12} e^{i\omega_{m_1,l_1}t + 2\eta.t}}{(i\omega_{m_1,l_1} + 2\eta)^2 + (\omega_{l_2,n_2})^2}\Big[(i\omega_{m_1,l_1} + 2\eta - j\omega_{l_2,n_2})(\cos(\omega_{l_2,n_2}t) + j\sin(\omega_{l_2,n_2}t)\Big]\left(\frac{1}{j\omega_{l_2,n_2} + \eta}\right).$$

If we now use again the Euler formula we have

$$I_{12}(t) = \frac{\beta_{12} i e^{i\omega_{m_1,l_1}t + 2\eta.t}}{(i\omega_{m_1,l_1} + 2\eta)^2 + (\omega_{l_2,n_2})^2}\Big[(i\omega_{m_1,l_1} + 2\eta - j\omega_{l_2,n_2}) e^{j\omega_{l_2,n_2}t}\Big]\left(\frac{j}{j\omega_{l_2,n_2} + \eta}\right)$$

or

$$I_{12}(t) = \frac{\beta_{12} e^{i\omega_{m_1,l_1}t + 2\eta.t}}{(i\omega_{m_1,l_1} + 2\eta)^2 + (\omega_{l_2,n_2})^2}\Big[(i\omega_{m_1,l_1} + 2\eta - j\omega_{l_2,n_2})\Big]\left(\frac{e^{j\omega_{l_2,n_2}t}}{j\omega_{l_2,n_2} + \eta}\right).$$

Now, note that the denominator could be written in a more symmetric form as

$$D = \frac{1}{(i\omega_{m_1,l_1} + 2\eta)^2 + (\omega_{l_2,n_2})^2} = \frac{1}{(i\omega_{m_1,l_1} + 2\eta + j\omega_{l_2,n_2})(i\omega_{m_1,l_1} + 2\eta - j\omega_{l_2,n_2})}$$

or

$$D = \frac{1}{(i\omega_{m_1,l_1} + 2\eta + j\omega_{l_2,n_2})} \cdot \frac{1}{(i\omega_{m_1,l_1} + 2\eta - j\omega_{l_2,n_2})}.$$

Hence, we can simplify the expression of the $I_{12}(t)$ integral in the form

$$I_{12}(t)=\beta_{12}e^{i\omega_{m_1,l_1}t+2\eta.t}\left[\frac{1}{(i\omega_{m_1,l_1}+2\eta+j\omega_{l_2,n_2})}\cdot\frac{1}{(i\omega_{m_1,l_1}+2\eta-j\omega_{l_2,n_2})}\right]\cdot$$
$$\cdot\qquad\left[i\omega_{m_1,l_1}+2\eta-j\omega_{l_2,n_2}\right]\left(\frac{e^{j\omega_{l_2,n_2}t}}{j\omega_{l_2,n_2}+\eta}\right)$$

or

$$I_{12}(t)=\beta_{12}e^{i\omega_{m_1,l_1}t+2\eta.t}\left[\frac{1}{(i\omega_{m_1,l_1}+j\omega_{l_2,n_2}+2\eta)}\right]\left(\frac{e^{j\omega_{l_2,n_2}t}}{j\omega_{l_2,n_2}+\eta}\right).$$

Using the fact that $q^{-1}=\overline{q}/(q\overline{q})$, we can write the result as

$$I_{12}(t)=\beta_{12}e^{i\omega_{m_1,l_1}t+2\eta.t}\cdot\frac{(-i\omega_{m_1,l_1}-j\omega_{l_2,n_2}+2\eta)}{\left((\omega_{m_1,l_1})^2+(\omega_{l_2,n_2})^2+4\eta^2\right)}\cdot\frac{(-j\omega_{l_2,n_2}+\eta)}{\left((\omega_{l_2,n_2})^2+\eta^2\right)}\cdot e^{j\omega_{l_2,n_2}t}.$$

By applying the same reasoning, one can get for the first and third integrals the expressions

$$I_{11}(t)=\beta_{11}e^{i\omega_{m_1,l_1}t+2\eta.t}\left[\frac{1}{(i\omega_{m_1,l_1}+i\omega_{l_1,n_1}+2\eta)}\right]\left(\frac{e^{i\omega_{l_1,n_1}t}}{i\omega_{l_1,n_1}+\eta}\right)$$

or

$$I_{11}(t)=\beta_{11}e^{i\omega_{m_1,l_1}t+2\eta.t}\cdot\frac{-i(\omega_{m_1,l_1}+\omega_{l_1,n_1})+2\eta}{\left((\omega_{m_1,l_1}+\omega_{l_1,n_1})^2+4\eta^2)\right)}\cdot\frac{(-i\omega_{l_1,n_1}+\eta)}{\left((\omega_{l_1,n_1})^2+\eta^2\right)}\cdot e^{i\omega_{l_1,n_1}t},$$

And

$$I_{13}(t)=\beta_{13}e^{i\omega_{m_1,l_1}t+2\eta.t}\left[\frac{1}{(i\omega_{m_1,l_1}+k\omega_{l_3,n_3}+2\eta)}\right]\left(\frac{e^{k\omega_{l_3,n_3}t}}{k\omega_{l_3,n_3}+\eta}\right)$$

or

$$I_{13}(t)=\beta_{13}e^{i\omega_{m_1,l_1}t+2\eta.t}\cdot\frac{(-i\omega_{m_1,l_1}-k\omega_{l_3,n_3}+2\eta)}{\left((\omega_{m_1,l_1})^2+(\omega_{l_3,n_3})^2+4\eta^2\right)}\cdot\frac{(-k\omega_{l_3,n_3}+\eta)}{\left((\omega_{l_3,n_3})^2+\eta^2\right)}\cdot e^{k\omega_{l_3,n_3}t}.$$

So, in general one can write

$$I_{xy}(t) = \beta_{xy} e^{q_x \omega_{m_x,l_x} t + 2\eta . t} \cdot \frac{(-q_x \omega_{m_x,l_x} - q_y \omega_{l_y,n_y} + 2\eta)}{\left((\omega_{m_x,l_x})^2 + (\omega_{l_y,n_y})^2 + 4\eta^2\right)} \cdot \frac{(-q_y \omega_{l_y,n_y} + \eta)}{\left((\omega_{l_y,n_y})^2 + \eta^2\right)} \cdot e^{q_y \omega_{l_y,n_y} t} ,$$

or

$$I_{xy}(t) = \delta_{xy} \beta_{xy} e^{q_x \omega_{m_x,l_x} t + 2\eta . t} \cdot \frac{(-q_x \omega_{m_x,l_x} - q_y \omega_{l_y,n_y} + 2\eta)}{\left((\omega_{m_x,l_x} + \omega_{l_y,n_y})^2 + 4\eta^2\right)} \cdot \frac{(-q_y \omega_{l_y,n_y} + \eta)}{\left((\omega_{l_y,n_y})^2 + \eta^2\right)} \cdot e^{q_y \omega_{l_y,n_y} t} ,$$

where $q_x$ ($q_y$) are imaginary units such that for $x$ (or $y$) = [1,2,3], we have [ $q_1 = i$, $q_2 = j$ , $q_3 = k$] and in the second equation, $\delta_{xy} = 1$ if $x = y$ and 0 otherwise .

# Appendix 4: Calculating the transition probability

To obtain the transition probability between two states it is necessary to calculate the modulus of the corresponding transition coefficient. In practical terms, we then need to calculate $I_{mn}.\bar{I}_{pq}$, where *m, n, p, q* = 1, 2 or 3.

However, it is worthwhile first to point out that if we expand the complex exponentials in the integral $I_2$, for example, in terms of sine and cosine functions, each integral gives rise to 24 terms. Hence, the product of one integral by the conjugate of another one will involve $(24)^2$ terms. To avoid this hard work we will proceed in a different way by proving that any of the ($m = p = n \neq q$, $m = p \neq n \neq q$, $m = n \neq p \neq q$ or any variation of these three situations) crossing terms will give a vanishing contribution, so that only non-crossing terms ($m = p = n = q$, $m = p \neq n = q$ or any variation of these two situations) will actually contribute to the process.

***Case 1: m = p ≠ n = q.***

For $m = 2$ and $n = 1$, for example, $I_{12}$ and its complex conjugate are given by

$$I_{12}(t) = \beta e^{2\eta.t} e^{i\omega_{m_1,l_1}t} \cdot \frac{(-i\omega_{m_1,l_1} - j\omega_{l_2,n_2} + 2\eta)}{\left((\omega_{m_1,l_1})^2 + (\omega_{l_2,n_2})^2 + 4\eta^2\right)} \cdot \frac{(-j\omega_{l_2,n_2} + \eta)}{\left((\omega_{l_2,n_2})^2 + \eta^2\right)} \cdot e^{j\omega_{l_2,n_2}t}$$

and

$$\bar{I}_{12}(t) = \beta e^{2\eta.t} e^{-j\omega_{l_2,n_2}t} \cdot \frac{(+j\omega_{l_2,n_2} + \eta)}{\left((\omega_{l_2,n_2})^2 + \eta^2\right)} \cdot \frac{(i\omega_{m_1,l_1} + j\omega_{l_2,n_2} + 2\eta)}{\left((\omega_{m_1,l_1})^2 + (\omega_{l_2,n_2})^2 + 4\eta^2\right)} \cdot e^{-i\omega_{m_1,l_1}t} \quad .$$

Multiplying both expressions we get

$$I_{12}(t)\bar{I}_{12}(t) = \beta^2 e^{4\eta.t} \cdot \frac{(\omega_{l_2,n_2})^2 + \eta^2}{\left((\omega_{l_2,n_2})^2 + \eta^2\right)^2} \cdot \frac{(\omega_{m_1,l_1})^2 + (\omega_{l_2,n_2})^2 + 4\eta^2}{\left((\omega_{m_1,l_1})^2 + (\omega_{l_2,n_2})^2 + 4\eta^2\right)^2}$$

and, finally,

$$I_{12}(t)\overline{I}_{12}(t)=\overline{I}_{12}(t)I_{12}(t)=\left|I_{12}(t)\right|^2=\beta^2 e^{4\eta .t}\cdot\frac{1}{\left((\omega_{l_2,n_2})^2+\eta^2\right)}\cdot\frac{1}{\left((\omega_{m_1,l_1})^2+(\omega_{l_2,n_2})^2+4\eta^2\right)}\quad .$$

***Case 2: m = p ≠ n ≠ q.***

Let's make $m = p = 1$, $n = 2$ and $q = 3$, for example, and calculate the term $I_{12}(t).\overline{I}_{13}(t)$.

In this case we have

$$I_{12}(t)=\beta_{12}e^{2\eta .t}e^{i\omega_{m_1,l_1}t}\cdot\frac{(-i\omega_{m_1,l_1}-j\omega_{l_2,n_2}+2\eta)}{\left((\omega_{m_1,l_1})^2+(\omega_{l_2,n_2})^2+4\eta^2\right)}\cdot\frac{(-j\omega_{l_2,n_2}+\eta)}{\left((\omega_{l_2,n_2})^2+\eta^2\right)}.e^{j\omega_{l_2,n_2}t}$$

For $I_3$, we have

$$\overline{I}_{13}(t)=\beta_{13}e^{2\eta .t}e^{-k\omega_{l_3,n_3}t}\cdot\frac{(k\omega_{l_3,n_3}+\eta)}{\left((\omega_{l_3,n_3})^2+\eta^2\right)}\cdot\frac{(i\omega_{m_1,l_1}+k\omega_{l_3,n_3}+2\eta)}{\left((\omega_{m_1,l_1})^2+(\omega_{l_3,n_3})^2+4\eta^2\right)}\cdot e^{-i\omega_{m_1,l_1}t}$$

Multiplying both expressions, we get

$$\begin{aligned}I_{12}(t)\overline{I}_{13}(t)=\frac{\beta_{12}\beta_{13}e^{4\eta .t}}{D_{12}.D_{13}}\Big[&(e^{i\omega_{m_1,l_1}t})(-i\omega_{m_1,l_1}-j\omega_{l_2,n_2}+2\eta)(-j\omega_{l_2,n_2}+\eta)\cdot\\&\cdot[\cos(\omega_{l_2,n_2}t)+j\sin(\omega_{l_2,n_2}t)][\cos(\omega_{l_3,n_3}t)-k\sin(\omega_{l_3,n_3}t)]\cdot\\&\cdot(k\omega_{l_3,n_3}+\eta)(i\omega_{m_1,l_1}+k\omega_{l_3,n_3}+2\eta)(e^{-i\omega_{m_1,l_1}t})\Big]\end{aligned}\quad ,$$

where we used

$$D_{13}=\left((\omega_{l_3,n_3})+\eta^2\right)\left((\omega_{m_1,l_1})^2+(\omega_{l_3,n_3})^2+4\eta^2\right)$$

and

$$D_{12}=\left((\omega_{m_1,l_1})^2+(\omega_{l_2,n_2})^2+4\eta^2\right)\left((\omega_{l_2,n_2})^2+\eta^2\right).$$

Using $jk = i$ and rearranging terms we get

$$I_{12}(t)\bar{I}_{13}(t)=\frac{\beta_{12}\beta_{13}ie^{4\eta.t}}{D_{12}.D_{13}}\begin{bmatrix}(e^{i\omega_{m_1,l_1}t})[(2\eta^2-\omega^2_{l_2,n_2}-i\eta\omega_{m_1,l_1})-j(3\eta\omega_{l_2,n_2}+i\omega_{m_1,l_1}\omega_{l_2,n_2})]\times\\ [\cos(\omega_{l_2,n_2}t)\cos(\omega_{l_3,n_3}t)-i\sin(\omega_{l_2,n_2}t)\sin(\omega_{l_3,n_3}t)+\\ j(\sin(\omega_{l_2,n_2}t)\cos(\omega_{l_3,n_3}t)+i\cos(\omega_{l_2,n_2}t)\sin(\omega_{l_3,n_3}t))]\times\\ [(2\eta^2-\omega^2_{l_3,n_3}+i\eta\omega_{m_1,l_1})-j(\omega_{m_1,l_1}\omega_{l_3,n_3}-i3\eta\omega_{l_3,n_3})](e^{-i\omega_{m_1,l_1}t})\end{bmatrix}.$$

Now, let´s define the complex quantities $C_n$ where $n = 1, 2, 3, 4, 5$ and $6$ as

$$\begin{aligned}\bar{C}_1&=2\eta^2-\omega^2_{l_2,n_2}-i\eta\omega_{m_1,l_1}\\ C_2&=3\eta\omega_{l_2,n_2}+i\omega_{m_1,l_1}\omega_{l_2,n_2}\\ \bar{C}_3(t)&=\cos(\omega_{l_2,n_2}t)\cos(\omega_{l_3,n_3}t)-i\sin(\omega_{l_2,n_2}t)\sin(\omega_{l_3,n_3}t)\\ C_4(t)&=\sin(\omega_{l_2,n_2}t)\cos(\omega_{l_3,n_3}t)+i\cos(\omega_{l_2,n_2}t)\sin(\omega_{l_3,n_3}t)\\ C_5&=2\eta^2-\omega^2_{l_3,n_3}+i\eta\omega_{m_1,l_1}\\ \bar{C}_6&=\omega_{m_1,l_1}\omega_{l_3,n_3}-i3\eta\omega_{l_3,n_3}\quad .\end{aligned}$$

With these C-definitions, we can rewrite the above expression as

$$I_{12}(t)\bar{I}_{13}(t)=\frac{\beta_{12}\beta_{13}e^{4\eta.t}}{D_{12}.D_{13}}\left[(e^{i\omega_{m_1,l_1}t})[\bar{C}_1-jC_2][\bar{C}_3+jC_4][C_5-j\bar{C}_6](e^{-i\omega_{m_1,l_1}t})\right]$$

or

$$I_{12}(t)\bar{I}_{13}(t)=\frac{\beta_{12}\beta_{13}e^{4\eta.t}}{D_{12}.D_{13}}\begin{bmatrix}[\bar{C}_1\bar{C}_4\bar{C}_6-\bar{C}_2C_3\bar{C}_6+\bar{C}_1\bar{C}_3C_5+\bar{C}_2C_4C_5]+\\ +j[C_1C_4C_5-C_2\bar{C}_3C_5-C_1C_3\bar{C}_6-C_2\bar{C}_4\bar{C}_6]e^{-i2\omega_{m_1,l_1}t}\end{bmatrix}.$$

Let´s proceed in the same way and calculate $I_{13}(t).\bar{I}_{12}(t)$. For $I_{12}$ and $I_{13}$, we have

$$\bar{I}_{12}(t)=\beta_{12}e^{2\eta.t}\frac{[\cos(\omega_{l_2,n_2}t)-j\sin(\omega_{l_2,n_2}t)](j\omega_{l_2,n_2}+\eta)(i\omega_{m_1,l_1}+j\omega_{l_2,n_2}+2\eta)(e^{-i\omega_{m_1,l_1}t})}{\left((\omega_{l_2,n_2})^2+\eta^2\right)\left((\omega_{m_1,l_1})^2+(\omega_{l_2,n_2})^2+4\eta^2\right)}$$

and

$$I_{13}(t)=\beta_{13}e^{2\eta.t}\cdot\frac{(e^{i\omega_{m_1,l_1}t})(-i\omega_{m_1,l_1}-k\omega_{l_3,n_3}+2\eta)(-k\omega_{l_3,n_3}+\eta)[\cos(\omega_{l_3,n_3}t)+k\sin(\omega_{l_3,n_3}t)]}{\left((\omega_{m_1,l_1})^2+(\omega_{l_3,n_3})^2+4\eta^2\right)\left((\omega_{l_3,n_3})^2+\eta^2\right)}.$$

Multiplying the above expressions and rewriting the expression in a sympletic form, we get

$$I_{13}(t)\bar{I}_{12}(t)=\frac{\beta_{12}\beta_{13}e^{4\eta.t}}{D_{12}.D_{13}}\Big\{(e^{i\omega_{m_1,l_1}t})[(2\eta^2-\omega_{l_3,n_3}^2-i\eta\omega_{m_1,l_1})-j(\omega_{l_3,n_3}\omega_{m_1,l_1}-i3\eta\omega_{l_3,n_3})]\cdot$$
$$\cdot[\cos(\omega_{l_3,n_3}t)\cos(\omega_{l_2,n_2}t)+i\sin(\omega_{l_3,n_3}t)\sin(\omega_{l_2,n_2}t)+$$
$$-j(\cos(\omega_{l_3,n_3}t)\sin(\omega_{l_2,n_2}t)-i\sin(\omega_{l_3,n_3}t)\cos(\omega_{l_2,n_2}t))]\cdot$$
$$\cdot[(2\eta^2-\omega_{l_2,n_2}^2+i\eta\omega_{m_1,l_1})+j(3\eta\omega_{l_2,n_2}+i\omega_{m_1,l_1}\omega_{l_2,n_2})](e^{-i\omega_{m_1,l_1}t}) \quad .$$

After defining similar complex quantities $C_n$

$$C_1=2\eta^2-\omega_{l_2,n_2}^2+i\eta\omega_{m_1,l_1}$$
$$C_2=3\eta\omega_{l_2,n_2}+i\omega_{m_1,l_1}\omega_{l_2,n_2}$$
$$C_3(t)=\cos(\omega_{l_2,n_2}t)\cos(\omega_{l_3,n_3}t)+i\sin(\omega_{l_2,n_2}t)\sin(\omega_{l_3,n_3}t)$$
$$C_4(t)=\sin(\omega_{l_2,n_2}t)\cos(\omega_{l_3,n_3}t)+i\cos(\omega_{l_2,n_2}t)\sin(\omega_{l_3,n_3}t)$$
$$\bar{C}_5=2\eta^2-\omega_{l_3,n_3}^2-i\eta\omega_{m_1,l_1}$$
$$\bar{C}_6=\omega_{m_1,l_1}\omega_{l_3,n_3}-i3\eta\omega_{l_3,n_3}$$

we get

$$I_{13}(t)\bar{I}_{12}(t)=\frac{\beta_{12}\beta_{13}e^{4\eta.t}}{D_{12}.D_{13}}\Big[(e^{i\omega_{m_1,l_1}t})[\bar{C}_5+j\bar{C}_6][C_3-jC_4][C_1+jC_2](e^{-i\omega_{m_1,l_1}t})\Big]$$

or

$$I_{13}(t)\bar{I}_{12}(t)=\frac{\beta_{12}\beta_{13}e^{4\eta.t}}{D_{12}.D_{13}}\Big\{[\bar{C}_5C_3C_1-C_6\bar{C}_3C_2+\bar{C}_5\bar{C}_4C_2+C_6C_4C_1]+$$
$$+j[C_5\bar{C}_3C_2+\bar{C}_6\bar{C}_4C_2+\bar{C}_6C_3C_1-C_5C_4C_1]e^{-i2\omega_{m_1,l_1}t}\Big\} \quad .$$

Since all $C$'s are complex quantities with the same imaginary unit ($i$, in this case), they commute among themselves. Let's re-arrange them and add $I_{12}(t).\bar{I}_{13}(t)$. Doing this, we get

$$I_{12}(t)\bar{I}_{13}(t)+I_{13}(t)\bar{I}_{12}(t)=\frac{\beta\lambda e^{4\eta.t}}{D_{12}.D_{13}}\Big\{[C_1C_4C_6-C_2\bar{C}_3C_6+C_1C_3\bar{C}_5+C_2\bar{C}_4\bar{C}_5]+$$
$$+j[C_2\bar{C}_3C_5-C_1C_4C_5+\bar{C}_1C_3C_6+C_2\bar{C}_4\bar{C}_6]e^{-i2\omega_{m_1,l_1}t}+$$
$$+[\bar{C}_1\bar{C}_4\bar{C}_6-\bar{C}_2C_3\bar{C}_6+\bar{C}_1\bar{C}_3C_5+\bar{C}_2C_4C_5]+$$
$$-j[C_2\bar{C}_3C_5-C_1C_4C_5+\bar{C}_1C_3C_6+C_2\bar{C}_4\bar{C}_6]e^{-i2\omega_{m_1,l_1}t}\Big\} \quad .$$

In the above expression one can see that the overall contribution of the *j*-terms is null. The same thing occurs with the real part of the complex numbers $Z_{nml} = C_n C_m C_l$, where $n, m, l = 1, 2, 3, 4, 5$ or $6$ and with any other $C$ conjugate that may appear in $Z$. With these considerations, the expression becomes a sum of purely real parts, i.e.,

$$I_{12}(t)\bar{I}_{13}(t) + I_{13}(t)\bar{I}_{12}(t) = \frac{\beta_{12}\beta_{13}e^{4\eta.t}}{D_{12}.D_{13}}\left\{\mathrm{Re}[C_1C_4C_6 - C_2\bar{C}_3C_6 + C_1C_3\bar{C}_5 + C_2\bar{C}_4\bar{C}_5] + \right.$$
$$\left. + \mathrm{Re}[\bar{C}_1\bar{C}_4\bar{C}_6 - \bar{C}_2C_3\bar{C}_6 + \bar{C}_1\bar{C}_3C_5 + \bar{C}_2C_4C_5]\right\} \quad .$$

But each one of the complex numbers that appear in the second bracket is the conjugate of a similar term in the first bracket, and therefore we can write

$$I_{12}(t)\bar{I}_{13}(t) + I_{13}(t)\bar{I}_{12}(t) = \frac{\beta_{12}\beta_{13}e^{4\eta.t}}{D_{12}.D_{13}}\left[2\,\mathrm{Re}[C_1C_4C_6 - C_2\bar{C}_3C_6 + C_1C_3\bar{C}_5 + C_2\bar{C}_4\bar{C}_5]\right] \quad .$$

As it should be expected, the modulus of a quaternionic number is a real quantity. The next step is to write explicitly the expressions for $R_nR_mR_l$ and analyze their properties. To do this let´s rewrite the Cs as canonical (i.e., $C_n = a_n + ib_n$) complex numbers and calculate explicitly $Re(C_nC_mC_l)$,

$$\begin{aligned}
C_1 &= 2\eta^2 - \omega_{l_2,n_2}^2 + i\eta\omega_{m_1,l_1} = a_1 + ib_1 \\
C_2 &= 3\eta\omega_{l_2,n_2} + i\omega_{m_1,l_1}\omega_{l_2,n_2} = a_2 + ib_2 \\
C_3(t) &= \cos(\omega_{l_2,n_2}t)\cos(\omega_{l_3,n_3}t) + i\sin(\omega_{l_2,n_2}t)\sin(\omega_{l_3,n_3}t) = a_3 + ib_3 \\
C_4(t) &= \sin(\omega_{l_2,n_2}t)\cos(\omega_{l_3,n_3}t) + i\cos(\omega_{l_2,n_2}t)\sin(\omega_{l_3,n_3}t) = a_4 + ib_4 \\
C_5 &= 2\eta^2 - \omega_{l_3,n_3}^2 + i\eta\omega_{m_1,l_1} = a_5 + ib_5 \\
C_6 &= \omega_{m_1,l_1}\omega_{l_3,n_3} + i3\eta\omega_{l_3,n_3} = a_6 + ib_6 \quad .
\end{aligned}$$

Hence,

$$C_nC_mC_l = (a_n + ib_n)(a_m + ib_m)(a_l + ib_l) \quad ,$$

$$C_nC_mC_l = (a_na_m - b_nb_m + ia_nb_m + ib_na_m)(a_l + ib_l)$$

$$C_nC_mC_l = (a_na_ma_l - b_nb_ma_l - a_nb_mb_l - b_na_mb_l) + i(a_nb_ma_l + b_na_ma_l + a_na_mb_l - b_nb_mb_l) \quad .$$

So, we can see that

$$\mathrm{Re}(C_1C_4C_6) = (a_1b_4a_6 + b_1a_4a_6 + a_1a_4b_6 - b_1b_4b_6)$$

$$\begin{aligned}\mathrm{Re}(C_1C_4C_6) = &-(\omega_{m_1,l_1}\omega_{l_3,n_3})(\omega_{l_2,n_2}^2 + \eta^2)\cos(\omega_{l_2,n_2}t)\sin(\omega_{l_3,n_3}t) + \\ &+\eta\omega_{l_3,n_3}(6\eta^2 + \omega_{m_1,l_1}^2 - 3\omega_{l_2,n_2}^2)\sin(\omega_{l_2,n_2}t)\cos(\omega_{l_3,n_3}t)\end{aligned}$$

$$\mathrm{Re}(C_2\bar{C}_3C_6) = (a_2a_3a_6 - b_2a_3b_6 + a_2b_3b_6 + b_2b_3a_6)$$

$$\mathrm{Re}(C_2\bar{C}_3C_6) = \omega_{l_2,n_2}\omega_{l_3,n_3}(9\eta^2 + \omega_{m_1,l_1}^2)\sin(\omega_{l_2,n_2}t)\sin(\omega_{l_3,n_3}t)$$

$$\mathrm{Re}(C_1C_3\bar{C}_5) = (a_1a_3a_5 - b_1b_3a_5 + a_1b_3b_5 + b_1a_3b_5)$$

$$\begin{aligned}\mathrm{Re}(C_1C_3\bar{C}_5) = &(4\eta^4 - \eta^2(2\omega_{l_2,n_2}^2 + 2\omega_{l_3,n_3}^2 + \omega_{m_1,l_1}^2) + \omega_{l_2,n_2}^2\omega_{l_3,n_3}^2)\cos(\omega_{l_2,n_2}t)\cos(\omega_{l_3,n_3}t) + \\ &+\eta\omega_{m_1,l_1}(\omega_{l_3,n_3}^2 - \omega_{l_2,n_2}^2)\sin(\omega_{l_2,n_2}t)\sin(\omega_{l_3,n_3}t) \quad ,\end{aligned}$$

and, finally,

$$\mathrm{Re}(C_2\bar{C}_4\bar{C}_5) = (a_2a_4a_5 - a_2b_4b_5 + b_2b_4a_5 + b_2a_4b_5)$$

$$\begin{aligned}\mathrm{Re}(C_2\bar{C}_4\bar{C}_5) = &\eta\omega_{l_2,n_2}(6\eta^2 + \omega_{m_1,l_1}^2 - 3\omega_{l_3,n_3}^2)\sin(\omega_{l_2,n_2}t)\cos(\omega_{l_3,n_3}t) + \\ &-\omega_{m_1,l_1}\omega_{l_2,n_2}(\eta^2 + \omega_{l_3,n_3}^2)\cos(\omega_{l_2,n_2}t)\sin(\omega_{l_3,n_3}t) \quad .\end{aligned}$$

$$\begin{aligned}\mathrm{Re}(C_1C_4C_6) = &-(\omega_{m_1,l_1}\omega_{l_3,n_3})(\omega_{l_2,n_2}^2 + \eta^2)\cos(\omega_{l_2,n_2}t)\sin(\omega_{l_3,n_3}t) + \\ &+\eta\omega_{l_3,n_3}(6\eta^2 + \omega_{m_1,l_1}^2 - 3\omega_{l_2,n_2}^2)\sin(\omega_{l_2,n_2}t)\cos(\omega_{l_3,n_3}t)\end{aligned}$$

According to standard second-order perturbation theory, the next step after taking the modulus is to consider the time derivative and to take the appropriate limit $\eta \to 0$. But, within a total of seventy two crossing (interference) terms, only two of them were explicitly developed in the above result. This corresponds to just a little portion of a more prohibitively complex result. However, we actually do not need to consider the complete development, since all of them involve $\sin(\omega t)$ and $\cos(\omega t)$ terms whose oscillatory behavior fluctuates strongly (almost randomly). Hence, in analogy with what happens in the treatment of fluctuations of chaotic light [4], one can expect that these interference terms give a vanishing average contribution.

In other words, the overall contribution of all crossing terms appearing in the modulus of the resulting integrals must vanish after time averaging them, so that the only actual contributions should come from the non-cross terms. We conclude that if we apply the above reasoning to the other crossing terms $I_{mn}(t)\overline{I}_{pq}(t)$ with at least one different subscript, we will obtain the same result: a sum of (almost randomly) oscillating functions in $\omega_{m_x,l_x}$ and vanishing contributions (in average) to the total modulus. The same reasoning can be applied to other cases (such as $m = p = n \neq q$ and $m = n \neq p \neq q$, for example); although we will not resolve them explicitly here, we can send the results via e-mail upon request.

# Appendix 5: Some Examples of Interesting Results Obtained by the Use of the Above Formalism

**Case 1:** Conductance of Benzene-di-thiol (a conjugated molecule):[5] the perfect agreement between the curves for the calculated conductance (black line) and the one where co-tunneling terms were artificially suppressed (red line) is strong evidence that the ballistic contribution is dominant in this aromatic system. Features such as the presence of a Coulomb blockade and the intermediary peak at $\sim 0.8V$ can be thoroughly explained in terms of the contribution of individual molecular orbitals.

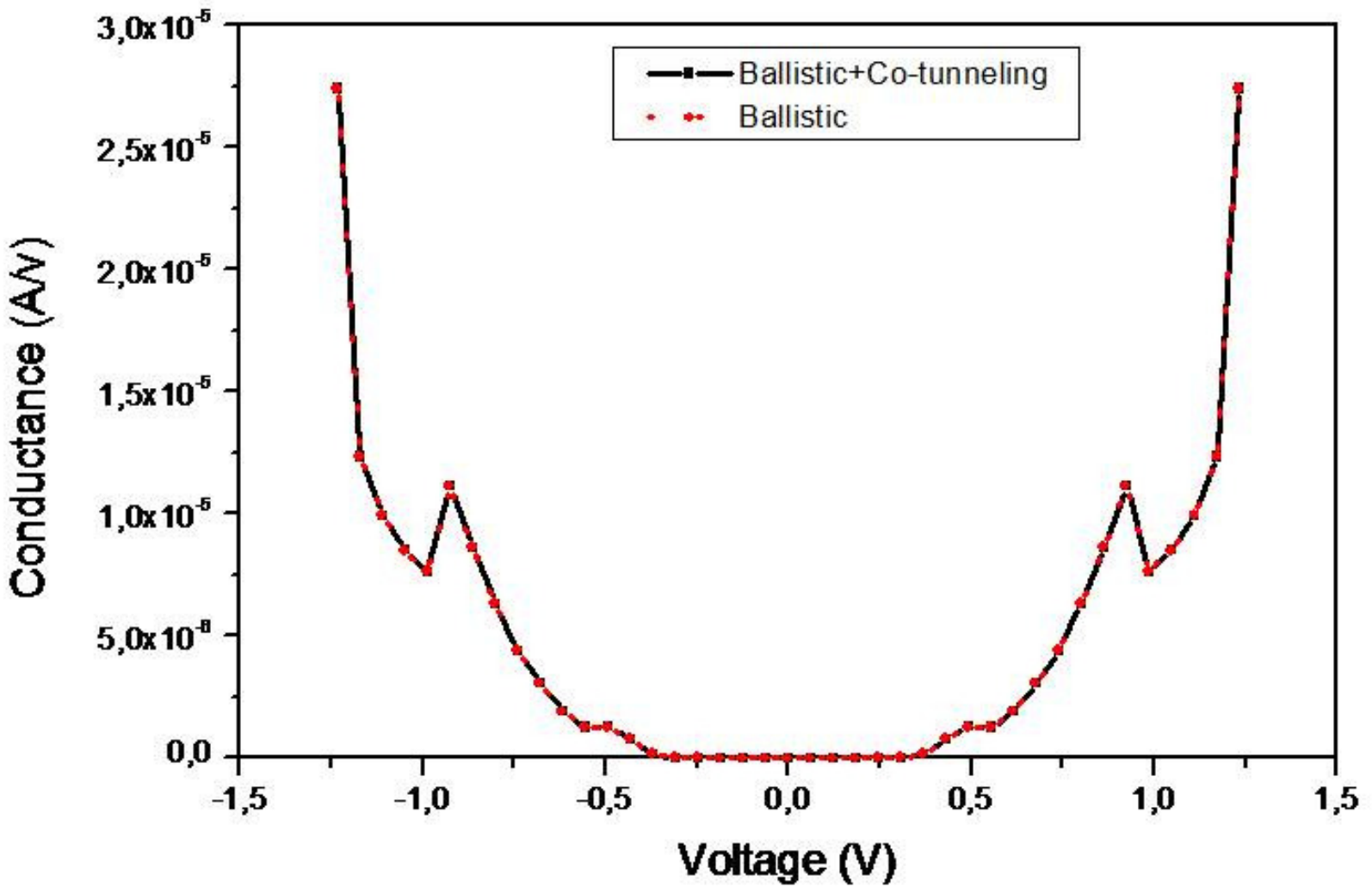

**Case 2:** Conductance of Octane-di-thiol (a saturated molecule):[6] in this case the ballistic contribution (red line) is negligible, so that the conductance is entirely dominated by non-coherent (i.e., co-tunneling) processes (black line). Features such as the presence of a Coulomb blockade and the intermediary peak at $\sim 0.8V$ can be thoroughly explained in terms of the contribution of individual molecular orbitals.

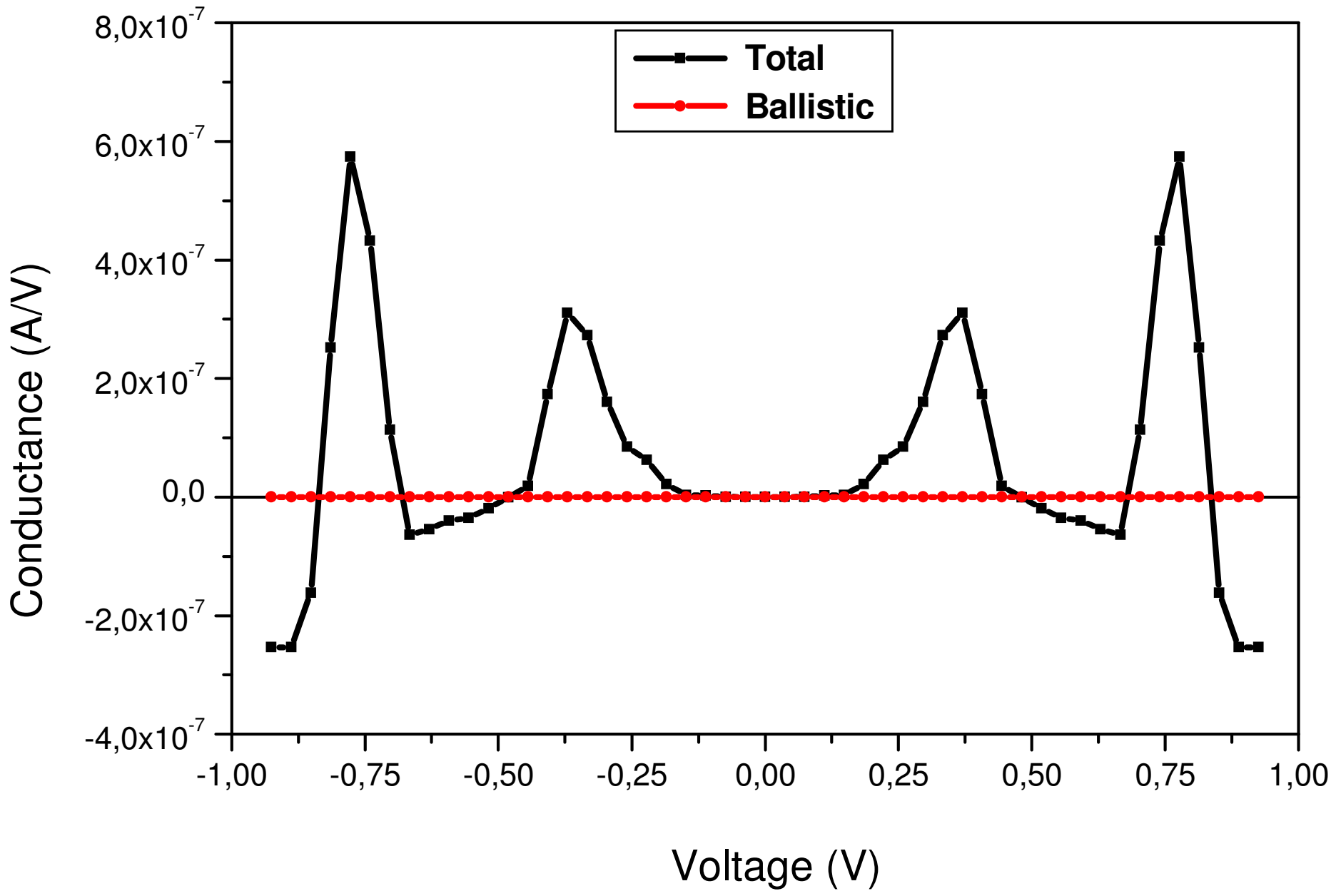

**Case 3:** Conductance of Lysine-di-thiol (intermediate case):[7] the comparison between the curves for the calculated conductance (black line) and the one where co-tunneling terms were artificially suppressed (red line) indicates that both ballistic and co-tunneling contributions are relevant in the transport process for this molecule. Specific features of these curves can be thoroughly explained in terms of the contribution of individual molecular orbitals.

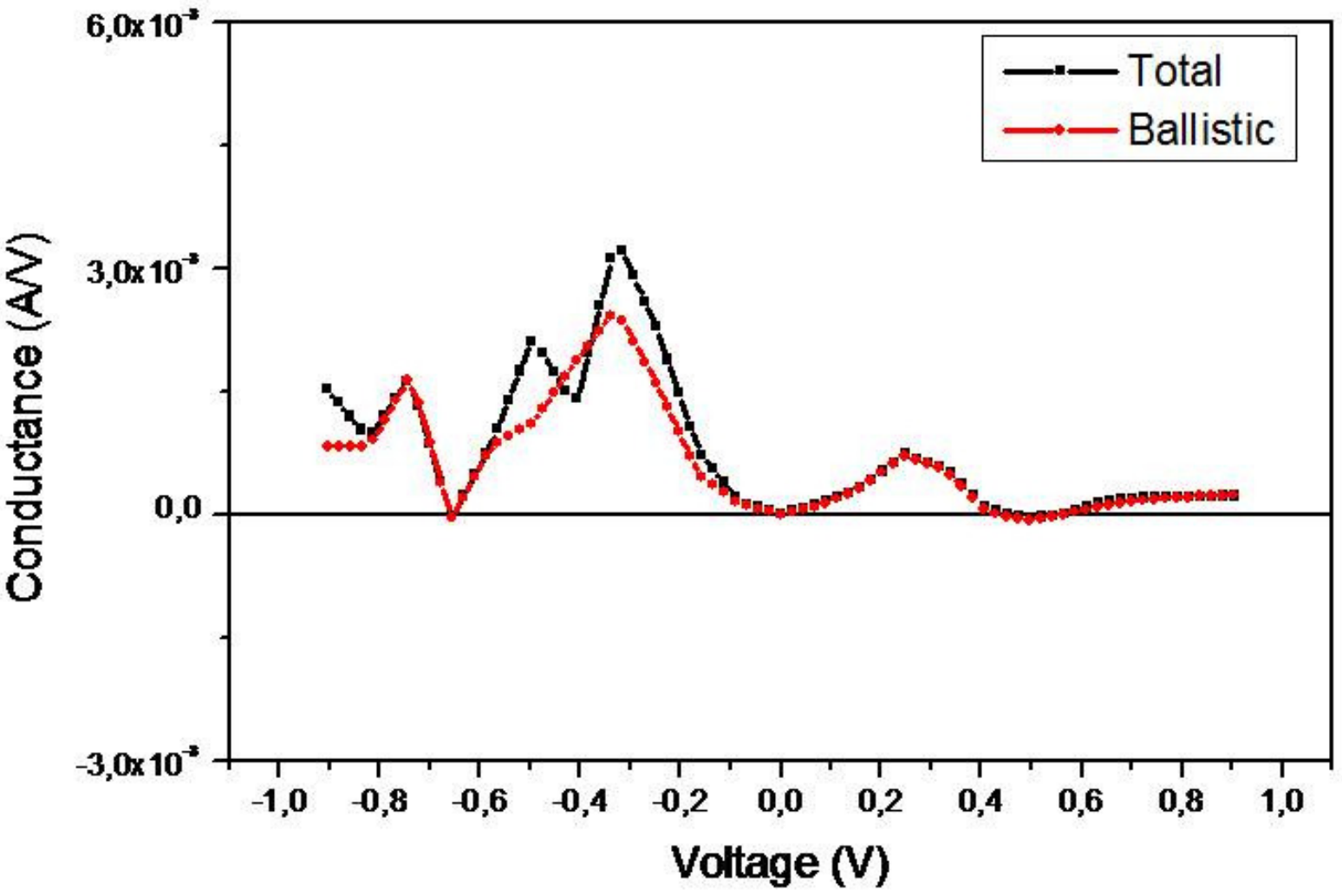

## Appendix 6: Time Ordering Operator in a Quaternionic System Time Dependent Perturbation Theory.

A well-known result in quantum mechanics is that related to the application of a time ordering operator ($P$) on the results of the time dependent perturbation theory (TDPT). $P$ is an operator that acts on the operators $A$ and $B$, for example, to order them so that the earliest time is to the right, i.e.:

$$P\left[A(t_1)B(t_2)\right]=\begin{cases}A(t_1)B(t_2) & t_1>t_2\\ B(t_2)A(t_1) & t_2>t_1\end{cases} .$$

For the second order TDPT case, the application of this operator gives

$$P\left(\int_{t_0}^{t} dt' T_I(t')\right)^2 = \int_{t_0}^{t} dt' \int_{t_0}^{t} dt'' P\left(T_I(t')T_I(t'')\right) = 2\int_{t_0}^{t} dt' \int_{t_0}^{t'} dt'' \left(T_I(t')T_I(t'')\right).$$

In a similar way, one can show that

$$P\left(\int_{t_0}^{t} dt' T_I(t')\right)^n = n!\int_{t_0}^{t} dt' \int_{t_0}^{t'} dt'' \ldots \int_{t_0}^{t^{n-1}} dt^n \left(T_I(t')T_I(t'')\ldots T_I(t^n)\right),$$

and, with this result, we can easily show that

$$C_k^{\infty}(t)=\left\langle k\right|P\sum_{l=0}^{\infty}\frac{1}{l!}\left(\frac{-i}{\hbar}\int_{t_0}^{t} dt' T_I(t')\right)^l\left|m\right\rangle=\left\langle k\right|P\exp\left(\frac{-i}{\hbar}\int_{t_0}^{t} dt' T_I(t')\right)\left|m\right\rangle .$$

In a similar way, we will show that by applying the same reasoning the time ordering operator acting in a quaternionic system will give the result

$$\gamma_1 C_{k_1}^{\infty}(t)=-i\gamma_1 P\exp\left(-\frac{i}{\hbar}\int_{t_0}^{t} dt' T_{1I}(t')-\frac{j}{\hbar}\int_{t_0}^{t} dt' T_{2I}(t')-\frac{k}{\hbar}\int_{t_0}^{t} dt' T_{3I}(t')\right) ,$$

with a similar expression for $c_{k_2}^{\infty}(t)$ and $c_{k_3}^{\infty}(t)$. Let´s start by rewriting Eq. (20) of the main text in a more compact form, considering explicitly the initial conditions $\left|C^{(0)}_{n_x}(0)\right| = 1$ for $x$ =1, 2 and 3, and noting that, without loss of generality, we can insert a constant phase factor like $C^{(0)}_{n_x}(0) = e^{\frac{-q_x\pi}{2}} = -q_x$ , where $q_1 = i$ , $q_2 = j$, $q_3 = k$. Doing this we will have

$$\gamma_1 C^{(2)}_{m_1}(t) = -i\gamma_1 \int_{t_0}^{t} dt' \int_{t_0}^{t'} dt'' \left[ \frac{1}{\hbar}\sum_{y=1}^{3}\sum_{n_y} \gamma_y^2 T_{m_y,l_y}(t') e^{q_y\omega_{m_y,l_y}t'} \left(\frac{-q_y}{\hbar}\right) \sum_{x=1}^{3}\sum_{n_x}\gamma_x^2 T_{l_x,n_x}(t'') e^{q_x\omega_{l_x,n_x}t''} C^{(0)}_{n_x}\right]$$

which, with $C^{(0)}_{n_x}$ as a purely imaginary number, can be rewritten as:

$$\gamma_1 C^{(2)}_{m_1}(t) = -i\gamma_1 \int_{t_0}^{t} dt' \int_{t_0}^{t'} dt'' \left[ \sum_{y=1}^{3}\sum_{n_y} \gamma_y^2 T_{m_y,l_y}(t') e^{q_y\omega_{m_y,l_y}t'} \left(\frac{-q_y}{\hbar}\right) \sum_{x=1}^{3}\sum_{n_x}\gamma_x^2 T_{l_x,n_x}(t'') e^{q_x\omega_{l_x,n_x}t''} \left(\frac{-q_x}{\hbar}\right)\right] .$$

Now note that although we have a sum over the sub-spaces (*x, y = 1, 2, 3*), the first sum depends on *t'* and the second one depends on *t"*. If we apply the time ordering operator in the above expression we obtain

$$P\left(\int_{t_0}^{t} dt' T_I(t')\right)^2 =$$

$$= -i\gamma_1 \int_{t_0}^{t} dt' \int_{t_0}^{t} dt'' P \left[ \sum_{y=1}^{3}\sum_{n_y} \gamma_y^2 T_{m_y,l_y}(t') e^{q_y\omega_{m_y,l_y}t'} \left(\frac{-q_y}{\hbar}\right) \sum_{x=1}^{3}\sum_{n_x}\gamma_x^2 T_{l_x,n_x}(t'') e^{q_x\omega_{l_x,n_x}t''} \left(\frac{-q_x}{\hbar}\right)\right]$$

and, hence,

$$P\left(\int_{t_0}^{t} dt' T_I(t')\right)^2 =$$

$$= -i\gamma_1 \int_{t_0}^{t} dt' \int_{t_0}^{t'} dt'' \left[ \sum_{y=1}^{3}\sum_{n_y} \gamma_y^2 T_{m_y,l_y}(t') e^{q_y\omega_{m_y,l_y}t'} \left(\frac{-q_y}{\hbar}\right) \sum_{x=1}^{3}\sum_{n_x}\gamma_x^2 T_{l_x,n_x}(t'') e^{q_x\omega_{l_x,n_x}t''} \left(\frac{-q_x}{\hbar}\right)\right] +$$

If we rewrite the second integral as $\int_{t_0}^{t} dt'' \int_{t_0}^{t''} dt'$, we have:

$$P\left(\int_{t_0}^{t} dt' T_I(t')\right)^2 =$$

$$= -i\gamma_1 \int_{t_0}^{t} dt' \int_{t_0}^{t'} dt'' \left[ \sum_{y=1}^{3} \sum_{n_y} \gamma_y^2 T_{m_y,l_y}(t') e^{q_y \omega_{m_y,l_y} t'} \left(\frac{-q_y}{\hbar}\right) \sum_{x=1}^{3} \sum_{n_x} \gamma_x^2 T_{l_x,n_x}(t'') e^{q_x \omega_{l_x,n_x} t''} \left(\frac{-q_x}{\hbar}\right) \right] +$$

$$-i\gamma_1 \int_{t_0}^{t} dt' \int_{t'}^{t} dt'' \left[ \sum_{y=1}^{3} \sum_{n_y} \gamma_y^2 T_{m_y,l_y}(t'') e^{q_y \omega_{m_y,l_y} t''} \left(\frac{-q_y}{\hbar}\right) \sum_{x=1}^{3} \sum_{n_x} \gamma_x^2 T_{l_x,n_x}(t') e^{q_x \omega_{l_x,n_x} t'} \left(\frac{-q_x}{\hbar}\right) \right]$$

$$= -i\gamma_1 \int_{t_0}^{t} dt' \int_{t_0}^{t'} dt'' \left[ \sum_{y=1}^{3} \sum_{n_y} \gamma_y^2 T_{m_y,l_y}(t') e^{q_y \omega_{m_y,l_y} t'} \left(\frac{-q_y}{\hbar}\right) \sum_{x=1}^{3} \sum_{n_x} \gamma_x^2 T_{l_x,n_x}(t'') e^{q_x \omega_{l_x,n_x} t''} \left(\frac{-q_x}{\hbar}\right) \right] +$$

$$-i\gamma_1 \int_{t_0}^{t} dt' \int_{t'}^{t} dt'' \left[ \sum_{y=1}^{3} \sum_{n_y} \gamma_y^2 T_{m_y,l_y}(t'') e^{q_y \omega_{m_y,l_y} t''} \left(\frac{-q_y}{\hbar}\right) \sum_{x=1}^{3} \sum_{n_x} \gamma_x^2 T_{l_x,n_x}(t') e^{q_x \omega_{l_x,n_x} t'} \left(\frac{-q_x}{\hbar}\right) \right]$$

$$= -i\gamma_1 \int_{t_0}^{t} dt' \int_{t_0}^{t'} dt'' \left[ \sum_{y=1}^{3} \sum_{n_y} \gamma_y^2 T_{m_y,l_y}(t') e^{q_y \omega_{m_y,l_y} t'} \left(\frac{-q_y}{\hbar}\right) \sum_{x=1}^{3} \sum_{n_x} \gamma_x^2 T_{l_x,n_x}(t'') e^{q_x \omega_{l_x,n_x} t''} \left(\frac{-q_x}{\hbar}\right) \right] +$$

$$-i\gamma_1 \int_{t_0}^{t} dt'' \int_{t_0}^{t''} dt' \left[ \sum_{y=1}^{3} \sum_{n_y} \gamma_y^2 T_{m_y,l_y}(t'') e^{q_y \omega_{m_y,l_y} t''} \left(\frac{-q_y}{\hbar}\right) \sum_{x=1}^{3} \sum_{n_x} \gamma_x^2 T_{l_x,n_x}(t') e^{q_x \omega_{l_x,n_x} t'} \left(\frac{-q_x}{\hbar}\right) \right]$$

$$= -i\gamma_1 \int_{t_0}^{t} dt' \int_{t_0}^{t'} dt'' \left[ \sum_{y=1}^{3} \sum_{n_y} \gamma_y^2 T_{m_y,l_y}(t') e^{q_y \omega_{m_y,l_y} t'} \left(\frac{-q_y}{\hbar}\right) \sum_{x=1}^{3} \sum_{n_x} \gamma_x^2 T_{l_x,n_x}(t'') e^{q_x \omega_{l_x,n_x} t''} \left(\frac{-q_x}{\hbar}\right) \right] +$$

$$-i\gamma_1 \int_{t_0}^{t} dt'' \int_{t_0}^{t''} dt' \left[ \sum_{y=1}^{3} \sum_{n_y} \gamma_y^2 T_{m_y,l_y}(t'') e^{q_y \omega_{m_y,l_y} t''} \left(\frac{-q_y}{\hbar}\right) \sum_{x=1}^{3} \sum_{n_x} \gamma_x^2 T_{l_x,n_x}(t') e^{q_x \omega_{l_x,n_x} t'} \left(\frac{-q_x}{\hbar}\right) \right]$$

and, upon interchanging *t'* and *t"*, the second term becomes equal to the first. Therefore we obtain:

$$P\left(\int_{t_0}^{t} dt' T_I(t')\right)^2 =$$

$$= -2i\gamma_1 \int_{t_0}^{t} dt' \int_{t_0}^{t'} dt'' \left[ \sum_{y=1}^{3} \sum_{n_y} \gamma_y^2 T_{m_y,l_y}(t') e^{q_y \omega_{m_y,l_y} t'} \left(\frac{-q_y}{\hbar}\right) \sum_{x=1}^{3} \sum_{n_x} \gamma_x^2 T_{l_x,n_x}(t'') e^{q_x \omega_{l_x,n_x} t''} \left(\frac{-q_x}{\hbar}\right) \right]$$

So, one can see that $P\left(\int_{t_0}^{t} dt' T_I(t')\right)^2 = 2C_{m_1}^{(2)}(t)$. In a similar way, we can show that

$$P\left(\int_{t_0}^{t} dt' T_I(t')\right)^n =$$

$$= n! C_{m_1}^{(n)}(t) = n!(-i\gamma_1)\int_{t_0}^{t} dt' \ldots \int_{t_0}^{t^{(m-1)}} dt^{(m)} \ldots \int_{t_0}^{t^{(n-1)}} dt^{(n)} \left[\sum_{y=1}^{3}\sum_{m_y} \gamma_y^2 T_{m_y,l_y}(t') e^{q_y \omega_{m_y,l_y} t'} \left(\frac{-q_y}{\hbar}\right) \ldots \right.$$

$$\left. \ldots \sum_{x=1}^{3}\sum_{r_x} \gamma_x^2 T_{r_x,s_x}\left(t^{(m)}\right) e^{q_x \omega_{r_x,s_x} t^{(m)}} \left(\frac{-q_x}{\hbar}\right) \ldots \sum_{z=1}^{3}\sum_{n_z} \gamma_z^2 T_{l_z,n_z}(t^{(n)}) e^{q_z \omega_{l_z,n_z} t^{(n)}} \left(\frac{-q_z}{\hbar}\right)\right]$$

and, as a consequence, after collecting all terms we obtain an exponential series, i.e.:

$$\gamma_1 C_{m_1}^{\infty}(t) = -i\gamma_1 P \exp\left(-\frac{i}{\hbar}\int_{t_0}^{t} dt' T_{1I}(t') - \frac{j}{\hbar}\int_{t_0}^{t} dt' T_{2I}(t') - \frac{k}{\hbar}\int_{t_0}^{t} dt' T_{3I}(t')\right) ,$$

where $T_{xI}(t') = \gamma_x^2 T_{m_x,l_x}(t') e^{q_x \omega_{m_x,l_x} t'}$ is the perturbation rewritten in the 'interaction picture' in such a way that $T_{xI}$ should not be viewed as a 'pure operator' but, instead, as matrix elements of the corresponding operator. This is due to the fact the $Q$ operator contains Hubbard-like operators in the off-diagonal positions, so that a summation over all possible states will not result in the identity matrix, as to be expected for the canonical projections operators in the diagonal positions in $Q$.

To rewrite the above expression in terms of operators, one can note that $T_{xI}(t') = \gamma_x^2 T_{m_x,l_x}(t') e^{q_x \omega_{m_x,l_x} t'}$ can be expressed as $\hat{T}_{xI}(t') = \gamma_x^2\left(e^{q_x H_x t'} T_x(t') e^{-q_x H_x t'}\right)$. So, if we rewrite Eq. (10) as

$$Q_\alpha(t) = |\chi_\alpha(t)\rangle\langle\chi_\alpha(t)| = \begin{pmatrix} \gamma_1\gamma_1 X_{11}(t) & \gamma_1\gamma_2 X_{12}(t) & \gamma_1\gamma_3 X_{13}(t) \\ \gamma_2\gamma_1 X_{21}(t) & \gamma_2\gamma_2 X_{22}(t) & \gamma_2\gamma_3 X_{23}(t) \\ \gamma_3\gamma_1 X_{31}(t) & \gamma_3\gamma_2 X_{32}(t) & \gamma_3\gamma_3 X_{33}(t) \end{pmatrix},$$

we obtain

$$Q_\alpha(t) T(t) = \begin{pmatrix} \gamma_1\gamma_1 X_{11}(t) T_1(t) & \gamma_1\gamma_2 X_{12}(t) T_2(t) & \gamma_1\gamma_3 X_{13}(t) T_3(t) \\ \gamma_2\gamma_1 X_{21}(t) T_1(t) & \gamma_2\gamma_2 X_{22}(t) T_2(t) & \gamma_2\gamma_3 X_{23}(t) T_3(t) \\ \gamma_3\gamma_1 X_{31}(t) T_1(t) & \gamma_3\gamma_2 X_{32}(t) T_2(t) & \gamma_3\gamma_3 X_{33}(t) T_3(t) \end{pmatrix} .$$

Hence, in the usual interaction representation picture for second order (for example) we have

$$\gamma_1 C_{m_1}^{(2)}(t) = -i\gamma_1 \int_{t_0}^{t} dt' \int_{t_0}^{t} dt'' \langle \varphi_{m_1} \| \left[ \sum_{y=1}^{3} \sum_{x=1}^{3} X_{1y}\gamma_y^2 e^{q_y H_y t'} T_y(t') e^{-q_y H_y t'} \left( \frac{-q_y}{\hbar} \right) \times \right.$$
$$\left. X_{yx}\gamma_x^2 e^{q_x H_x t''} T_x(t'') e^{-q_x H_x t''} \left( \frac{-q_x}{\hbar} \right) \right] \| \varphi_{n_x} \rangle$$

So that, finally, in terms of operators the time ordering process is given by

$$\gamma_1 U_I(t,t_0) = -i\gamma_1 P \exp\left( -\frac{i}{\hbar}\int_{t_0}^{t} dt' \left( X_{y1}\hat{T}_{1I}(t') \right) - \frac{j}{\hbar}\int_{t_0}^{t} dt' \left( X_{y2}\hat{T}_{2I}(t') \right) - \frac{k}{\hbar}\int_{t_0}^{t} dt' \left( X_{y3}\hat{T}_{3I}(t') \right) \right) .$$

Note that $X_{xy}(t')T_{zI}(t') = X_{xy}(t')T_{zI}(t')\delta_{yz}$ and that, for $x = y$, the $X_{xy}$ operator is nothing but the identity matrix when summing over all possible states. Also if $\gamma_1 = 1$ and $\gamma_2 = \gamma_3 = 0$ , for example, we recover the usual results for just one quaternionic sub-space.

From a practical point of view, the above results make a lot easier to deal with the non-commutative algebra of quaternions since we do not need to face the iterated integrals in the second order time dependent perturbation; rather, the above result permits us to deal with simple (non-iterative) integrals.

To be sure that the time ordering operator *P* give the same result as expressed by Eq. (25) of the main text, let´s write the (second order time ordered) integral called $I_{yx}^{(2)}$ in an explicit manner as

$$I_{yx}^{(2)} =$$

$$= -q_y\gamma_y \int_{-\infty}^{t} dt' \int_{-\infty}^{t} dt'' \left[ \sum_{l_y} \gamma_y^2 T_{m_y,l_y}(t') e^{q_y \omega_{m_y,l_y} t'} \left( \frac{-q_y}{\hbar} \right) \sum_{l_x} \gamma_x^2 T_{l_x,n_x}(t'') e^{q_x \omega_{l_x,n_x} t''} \left( \frac{-q_x}{\hbar} \right) \right]$$

$$= -q_y\gamma_y \int_{-\infty}^{t} dt' \left[ \sum_{l_y} \gamma_y^2 T_{m_y,l_y} e^{q_y \omega_{m_y,l_y} t' + \eta t'} \left( \frac{-q_y}{\hbar} \right) \right] \int_{-\infty}^{t} dt'' \left[ \sum_{l_x} \gamma_x^2 T_{l_x,n_x} e^{q_x \omega_{l_x,n_x} t'' + \eta t''} \left( \frac{-q_x}{\hbar} \right) \right]$$

$$= -q_y\gamma_y\left[\sum_{l_y}\frac{\gamma_y^2 T_{m_y,l_y} e^{q_y\omega_{m_y,l_y}t+\eta t}}{q_y\omega_{m_y,l_y}+\eta}\left(\frac{-q_y}{\hbar}\right)\sum_{l_x}\frac{\gamma_x^2 T_{l_x,n_x} e^{q_x\omega_{l_x,n_x}t+\eta t}}{q_x\omega_{l_x,n_x}+\eta}\left(\frac{-q_x}{\hbar}\right)\right]$$

Thus, the modulus of the above term can be expressed as

$$I_{yx}^{(2)} = -q_y\gamma_y\left[\sum_{l_y}\frac{\gamma_y^2 T_{m_y,l_y} e^{q_y\omega_{m_y,l_y}t+\eta t}}{q_y\omega_{m_y,l_y}+\eta}\left(\frac{-q_y}{\hbar}\right)\sum_{l_x}\frac{\gamma_x^2 T_{l_x,n_x} e^{q_x\omega_{l_x,n_x}t+\eta t}}{q_x\omega_{l_x,n_x}+\eta}\left(\frac{-q_x}{\hbar}\right)\right]$$

or

$$\left|I_{yx}^{(2)}\right|^2 = \gamma_y^2\frac{(\gamma_y^2\gamma_x^2)e^{4\eta t}}{\hbar^4}\left[\sum_{l_y}\sum_{l_x}\frac{\left|T_{m_y,l_y}\right|^2\left|T_{l_x,n_x}\right|^2}{\left(\omega_{m_y,l_y}^2+\eta^2\right)\left(\omega_{l_x,n_x}^2+\eta^2\right)}\right].$$

Now, deriving in time and taking the limit *η→0*, i.e., expanding the exponential as before, we have

$$\frac{d\left|I_{yx}^{(2)}\right|^2}{dt} = \gamma_y^2\frac{\left(\gamma_y^2\gamma_x^2\right)^2}{\hbar^4}\sum_{l_y}\sum_{l_x}\left|T_{m_y,l_y}\right|^2\left|T_{l_x,n_x}\right|^2\frac{(4\eta+16\eta^2 t)}{\left(\omega_{m_y,l_y}^2+\eta^2\right)\left(\omega_{l_x,n_x}^2+\eta^2\right)} \quad ,$$

and so, by using the definition of the Dirac delta one can write

$$\frac{d\left|I_{yx}^{(2)}\right|^2}{dt} =$$

$$= \gamma_y^2\frac{2\left(\gamma_y^2\gamma_x^2\right)^2}{\hbar^4}\sum_{l_y}\sum_{l_x}\left|T_{m_y,l_y}\right|^2\left|T_{l_x,n_x}\right|^2\left[\frac{2\delta\left(\omega_{l_x,n_x}\right)}{\omega_{m_y,l_y}^2}+8t\delta\left(\omega_{m_y,l_y}\right)\delta\left(\omega_{l_x,n_x}\right)\right]$$

$$= 2\gamma_y^2\frac{\left(\gamma_y^2\gamma_x^2\right)^2}{\hbar^2}\sum_{l_y}\sum_{l_x}\left|T_{m_y,l_y}\right|^2\left|T_{l_x,n_x}\right|^2\left[\frac{2\pi\hbar}{\left(E_{m_y}-E_{l_y}\right)^2}+8\pi^2 t\delta\left(E_{m_y}-E_{l_y}\right)\right]\delta\left(E_{l_x}-E_{n_x}\right) \quad ,$$

which is twice the result obtained in Eqs. (24) and (25) with $y$ = 1 and $x$ = 1, 2 and 3, i.e., all terms of order $O_{(0)}$ and $O_{(t)}$ that appear in those equations. Note that the Dirac delta "imposes" automatically that

$$\frac{\delta\left(E_{l_x}-E_{n_x}\right)}{\left(E_{m_y}-E_{l_y}\right)^2}=\frac{\delta\left(E_{l_x}-E_{n_x}\right)}{\left(E_{m_y}-E_{l_y}\right)^2+\left(E_{l_x}-E_{n_x}\right)^2} \quad .$$

# Appendix 7: Dyson Series and Green´s Function in a Quaternionic System.

The time ordering operator is a key tool in making a link between time dependent perturbation and the Dyson series and, consequently, for the construction of a series in terms of a generalized propagator or Green´s function that mixes quaternionic sub-spaces. In case of iterative integrals, for example, a change from one quaternionic subspace to another implies in dealing with a different complex unit that does not commute with the first. For this reason, it is not straightforward to define a propagator in these cases, for example when on going from *t'* to *t"* implies the change from *x*-subspace to *y*-subspace, where $x \neq y$.

However, this procedure can be easily done by recurring to the time ordering operator and, in fact, we can make the link (time dependent perturbation → Dyson series) by noting that

$$I_{yx}^{(2)} =$$

$$= -q_y\gamma_y \left[ \int_{-\infty}^{t} dt' \sum_{l_y} \gamma_y^2 T_{m_y,l_y} e^{q_y\omega_{m_y,l_y}t' + \eta t'} \left( \frac{-q_y}{\hbar} \right) \right] \left[ \int_{-\infty}^{t} dt'' \sum_{l_x} \gamma_x^2 T_{l_x,n_x} e^{q_x\omega_{l_x,n_x}t'' + \eta t''} \left( \frac{-q_x}{\hbar} \right) \right]$$

$$= -q_y\gamma_y \sum_{l_y} \gamma_y^2 T_{m_y,l_y} \left( \frac{-q_y}{\hbar} \right) \left[ \int_{-\infty}^{t} e^{q_y\omega_{m_y,l_y}t' + \eta t'} dt' \right] \sum_{l_x} \gamma_x^2 T_{l_x,n_x} \left( \frac{-q_x}{\hbar} \right) \left[ \int_{-\infty}^{t} e^{q_x\omega_{l_x,n_x}t'' + \eta t''} dt'' \right]$$

$$= -q_y\gamma_y \sum_{l_y} \gamma_y^2 T_{m_y,l_y} \left( \frac{-q_y}{\hbar} \right) \left[ \int_{-\infty}^{+\infty} e^{q_y\left(\frac{E_{m_y}}{\hbar}\right)t'} \theta(t-t') e^{-q_y\left(\frac{E_{l_y}+q_y\eta\hbar}{\hbar}\right)t'} dt' \right] \times$$

$$\sum_{l_x} \gamma_x^2 T_{l_x,n_x} \left( \frac{-q_x}{\hbar} \right) \left[ \int_{-\infty}^{+\infty} e^{q_x\left(\frac{-E_{n_x}}{\hbar}\right)t''} \theta(t-t'') e^{-q_x\left(\frac{-E_{l_x}+q_x\eta\hbar}{\hbar}\right)t''} dt'' \right] \quad ,$$

Now, introducing the new variables $t - t' = \tau_1$ and $t - t'' = \tau_2$ and rewriting $F(E_{n_x})$ as $F(E_{n_x}) = \int_{-\infty}^{+\infty} F(E)\delta(E - E_{n_x})dE$ , we have

$$I^{(2)}_{yx} =$$

$$-q_y\gamma_y\sum_{l_y}\gamma_y^2 T_{m_y,l_y}\left(\frac{-q_y}{\hbar}\right)\int_{-\infty}^{+\infty}\left[\int_{-\infty}^{+\infty} e^{q_y\left(\frac{-E_1}{\hbar}\right)(t-\tau_1)}\theta(\tau_1)e^{-q_y\left(\frac{E_{m_y}+q_y\eta\hbar}{\hbar}\right)(t-\tau_1)}d\tau_1\right]\delta(E_1-E_{l_y})dE_1\times$$

$$\sum_{l_x}\gamma_x^2 T_{l_x,n_x}\left(\frac{-q_x}{\hbar}\right)\int_{-\infty}^{+\infty}\left[\int_{-\infty}^{+\infty} e^{q_x\left(\frac{-E_2}{\hbar}\right)(t-\tau_2)}\theta(\tau_2)e^{-q_x\left(\frac{E_{l_x}+q_x\eta\hbar}{\hbar}\right)(t-\tau_2)}d\tau_2\right]\delta(E_2-E_{n_x})dE_2 \quad ,$$

$$I^{(2)}_{yx} = -q_y\gamma_y\sum_{l_y}\gamma_y^2 T_{m_y,l_y}\left(\frac{-q_y}{\hbar}\right)\times$$

$$\int_{-\infty}^{+\infty}\left[\int_{-\infty}^{+\infty} e^{q_y\left(\frac{-E_1}{\hbar}\right)(t-\tau_1)}\theta(\tau_1)e^{-q_y\left(\frac{E_{m_y}+q_y\eta\hbar}{\hbar}\right)(t-\tau_1)}d\tau_1\right]\delta(E_1-E_{l_y})dE_1\times$$

$$\sum_{l_x}\gamma_x^2 T_{l_x,n_x}\left(\frac{-q_x}{\hbar}\right)\int_{-\infty}^{+\infty}\left[\int_{-\infty}^{+\infty} e^{q_x\left(\frac{-E_2}{\hbar}\right)(t-\tau_2)}\theta(\tau_2)e^{-q_x\left(\frac{E_{l_x}+q_x\eta\hbar}{\hbar}\right)(t-\tau_2)}d\tau_2\right]\delta(E_2-E_{n_x})dE_2$$

which can be written as

$$I^{(2)}_{yx} = -q_y\gamma_y\sum_{l_y}\gamma_y^2 T_{m_y,l_y}\left(\frac{-q_y}{\hbar}\right)\times$$

$$\int_{-\infty}^{+\infty}\left[e^{-q_y\left(\frac{E_1-E_{m_y}+q_y\eta\hbar}{\hbar}\right)t}\cdot\int_{-\infty}^{+\infty} e^{q_y\left(\frac{E_1}{\hbar}\right)\tau_1}\theta(\tau_1)e^{-q_y\left(\frac{E_{m_y}-q_y\eta\hbar}{\hbar}\right)\tau_1}d\tau_1\right]\delta(E_1-E_{l_y})dE_1\times$$

$$\sum_{l_x}\gamma_x^2 T_{l_x,n_x}\left(\frac{-q_x}{\hbar}\right)\int_{-\infty}^{+\infty}\left[e^{-q_x\left(\frac{E_2-E_{l_x}+q_x\eta\hbar}{\hbar}\right)t}\cdot\int_{-\infty}^{+\infty} e^{q_x\left(\frac{E_2}{\hbar}\right)\tau_2}\theta(\tau_2)e^{-q_x\left(\frac{E_{l_x}-q_x\eta\hbar}{\hbar}\right)\tau_2}d\tau_2\right]\delta(E_2-E_{n_x})dE_2$$

Now note that the terms in brackets are nothing but the Fourier transform of the two functions

given by

$$G_x(\tau_2) = -q_x \hbar^{-1} \theta(\tau_2) \exp\left[-q_x(\hbar^{-1} E_{l_x} - q_x \eta)\tau_2\right]$$

and

$$G_y(\tau_1) = -q_y \hbar^{-1} \theta(\tau_1) \exp\left[-q_y(\hbar^{-1} E_{m_y} - q_y \eta)\tau_1\right] \ ,$$

so that one can rewrite the above equation as

$$I_{yx}^{(2)} = -q_y \frac{\gamma_y}{\hbar^2} e^{-2\eta t} \left(\gamma_y^2 \gamma_x^2\right) \phi_y(t) G_y^0(E_{m_y}) T_{m_y,l_y} \phi_x(t) G_x^0(E_{n_x}) T_{l_x,n_x} \ ,$$

where $G_x^0(E_{n_x}) = \left[\left(E_{n_x} - E_{l_x}\right)\hbar^{-1} + q_x \eta\right]^{-1}$ and $G_y^0(E_{l_y}) = \left[\left(E_{l_y} - E_{m_y}\right)\hbar^{-1} + q_y \eta\right]^{-1}$ are the Green´s function of the quaternionic sub-spaces *x* and *y*, respectively, and $\phi_y(t)$, $\phi_x(t)$ are phase factors given by $\phi_y(t) = \exp\left(q_y \omega_{m_y l_y} t\right)$ and $\phi_x(t) = \exp\left(q_x \omega_{l_x n_x} t\right)$. Note that for x = y all quantities in the above expression commute and the phase factor yields $\phi(t) = \phi_x(t)\phi_x(t) = \exp\left(q_x \omega_{m_x n_x} t\right)$ and do not affect the modulus when we are dealing with just one sub-space, i.e., without mixing quaternionic sub-spaces.

The previous equation can also be written in terms of operators. To do this let´s rewritten the Q operator again (see Eq. (10)) as

$$Q_\alpha(t) = |\chi_\alpha(t)\rangle\langle\chi_\alpha(t)| = \begin{pmatrix} \gamma_1\gamma_1 X_{11}(t) & \gamma_1\gamma_2 X_{12}(t) & \gamma_1\gamma_3 X_{13}(t) \\ \gamma_2\gamma_1 X_{21}(t) & \gamma_2\gamma_2 X_{22}(t) & \gamma_2\gamma_3 X_{23}(t) \\ \gamma_3\gamma_1 X_{31}(t) & \gamma_3\gamma_2 X_{32}(t) & \gamma_3\gamma_3 X_{33}(t) \end{pmatrix} .$$

Thus, in terms of *X*-operators we obtain

$$\hat{I}_{yx}^{(2)} = -q_z \frac{\gamma_z}{\hbar^2} X_{zy} \left(G_y^0(E_y)\hat{T}_{yI}(t)\right) X_{yx} \left(G_x^0(E_x)\hat{T}_{xI}(t)\right) \ ,$$

where the "hat" above $I_{xy}$ indicates that now we are dealing with an operator rather than with a matrix element and the phase ($\phi$) was incorporated in the $\boldsymbol{T}_{(t)}$ operator.

So, in general, one can write for high orders

$$\hat{I}^{(n)} = -q_y \frac{\gamma_y}{\hbar^n} \prod_{p=1}^{n} X_{y_p x_p} \left( G^0_{x_p}(E_{x_p}) \hat{T}_{I x_p}(t) \right) ,$$

where for each value of $p$, $x_p$ $(y_p) = 1, 2$ or $3$ labels the sub-spaces (like $y$). Note that if $x_p = y_p$, the $X$-operator becomes identical to the identity matrix when summed over all possible states.

*Corresponding author: celso@df.ufpe.br